%% file: main.tex
\documentclass[10pt,twocolumn,letterpaper]{article}

\usepackage[pagenumbers]{cvpr} 

\input{preamble}

\usepackage{multirow}
\usepackage{multicol}

\definecolor{cvprblue}{rgb}{0.21,0.49,0.74}
\usepackage[pagebackref,breaklinks,colorlinks,allcolors=cvprblue]{hyperref}
\usepackage[accsupp]{axessibility}

\def\paperID{13103} 
\def\confName{CVPR}
\def\confYear{2025}

\newcommand{\websiteURL}[0]{https://heheyas.github.io/MeshGen}
\newcommand{\websiteURLText}[0]{https://heheyas.github.io/MeshGen}
\newcommand{\codeURL}[0]{https://github.com/heheyas/MeshGen}

\newcommand{\websiteLink}[0]{{\color{blue}{\href{\websiteURL}{\texttt{\websiteURLText}}}}\xspace}

\newcommand{\approach}[0]{MeshGen\xspace}
\newcommand{\subtitle}[1]{\noindent\textbf{#1\xspace}}
\newcommand{\fref}[1]{Fig.~\ref{#1}\xspace}
\newcommand{\tref}[1]{Tab.~\ref{#1}\xspace}
\newcommand{\sref}[1]{Section.~\ref{#1}\xspace}
\newcommand{\aref}[1]{appendix~\ref{#1}\xspace}

\title{\approach: Generating PBR Textured Mesh with Render-Enhanced Auto-Encoder and Generative Data Augmentation}

\author{
Zilong Chen$^1$, Yikai Wang$^{2\dagger}$, Wenqiang Sun$^{3}$, Feng Wang$^{1}$, Yiwen Chen$^{4}$, Huaping Liu$^{1\dagger}$\\
\vspace{-1.5mm}
\small$^1$Beijing National Research Center for Information Science and Technology (BNRist),\\
\vspace{-1.5mm}
\small Department of Computer Science and Technology, Tsinghua University\\
\vspace{-1.5mm}
\small$^2$ School of Artificial Intelligence, Beijing Normal University \; $^3$ HKUST\\
\vspace{-1.5mm}
\small$^4$ S-Lab, School of Computer Science and Engineering, Nanyang Technological University\\
\vspace{-1.5mm}
{\tt\small chenzl22@mails.tsinghua.edu.cn, yikaiw@outlook.com, hpliu@tsinghua.edu.cn}\\
{\normalsize Project page: \websiteLink}
}

\begin{document}
\maketitle

\begin{abstract}
\let\thefootnote\relax\footnotetext{$^\dagger$Corresponding authors.}

\vspace{-5mm}
In this paper, we introduce \approach, an advanced image-to-3D pipeline that generates high-quality 3D meshes with detailed geometry and physically based rendering (PBR) textures. Addressing the challenges faced by existing 3D native diffusion models, such as suboptimal auto-encoder performance, limited controllability, poor generalization, and inconsistent image-based PBR texturing, \approach employs several key innovations to overcome these limitations. We pioneer a render-enhanced point-to-shape auto-encoder that compresses meshes into a compact latent space by designing perceptual optimization with ray-based regularization. This ensures that the 3D shapes are accurately represented and reconstructed to preserve geometric details within the latent space. To address data scarcity and image-shape misalignment, we further propose geometric augmentation and generative rendering augmentation techniques, which enhance the model's controllability and generalization ability, allowing it to perform well even with limited public datasets. For the texture generation, \approach employs a reference attention-based multi-view ControlNet for consistent appearance synthesis. This is further complemented by our multi-view PBR decomposer that estimates PBR components and a UV inpainter that fills invisible areas, ensuring a seamless and consistent texture across the 3D mesh. Our extensive experiments demonstrate that \approach largely outperforms previous methods in both shape and texture generation, setting a new standard for the quality of 3D meshes generated with PBR textures.
\end{abstract}

\section{Introduction}
With the rapid advancement of diffusion-based image generation models~\citep{stablediffusion, zhao2022egsde, zhao2025riflex, zhao2024identifying}, there has been significant progress in automatic 3D generation. In particular, methods utilizing score distillation sampling~\citep{dreamfusion} have demonstrated breakthroughs by leveraging priors from text-to-image diffusion models. However, these optimization-based methods are relatively slow and face challenges such as mode collapse~\citep{wang2023esd, prolificdreamer} and the Janus problem~\citep{perpneg, seo2023let} due to the lack of inherent 3D information.
Subsequent strategies tackle these challenges by focusing on multi-view generation~\citep{liu2023syncdreamer, long2023wonder3d, chen2024v3d, voleti2024sv3d} and large reconstruction models~\citep{zou2023triplane, tang2024lgm, LRM, PF-LRM, xu2024instantmesh, liu2024meshformer, wei2024meshlrm}. The former generates multi-views for 3D reconstruction, while the latter maps sparse-view images to compact neural 3D representations like triplane NeRF~\citep{Chan2021eg3d} or 3D Gaussians~\citep{zou2023triplane, tang2024lgm}. Although these methods enhance 3D generation quality and speed, they often use volumetric representations like NeRF or Gaussian instead of meshes, leading to quality loss during conversion~\citep{LRM, chen2024gsgen, tang2023dreamgaussian}. Additionally, relying solely on multi-views for supervision makes them prone to inconsistencies and challenges in reconstructing objects with complex geometry~\citep{sun2024freeplane}.

Recently, 3D native diffusion models have garnered significant attention as a promising paradigm toward mesh-oriented generation~\citep{gupta20233dgen, Wang2023rodin, zhang2024clay, li2024craftsman, wu2024direct3d, hong20243dtopia, chen2024primx}. By mapping meshes into a compact latent space using 3D auto-encoders, these methods directly learn the distribution of 3D shapes instead of reconstructing from generated multi-views. Despite considerable progress that has been made, several challenges remain unresolved.
Firstly, due to the lack of perceptual loss, existing 3D auto-encoders have limited capacity, making it difficult to capture rich surface details and high-frequency information.
Moreover, existing 3D native diffusion models typically generate simple and symmetric shapes that differ significantly from the reference image, and the scarcity and poor quality of public datasets further limit the generalization ability of current open-source models.
In addition, existing image-conditioned texture generation methods struggle to produce appearances consistent with the reference images and can only generate textures with light baked in, rather than the physically based rendering (PBR) materials required in practical applications.

In response to these challenges, we introduce \approach, a novel image-to-3D pipeline specially designed to generate PBR textured meshes that closely resemble the reference image in both geometry and appearance.
Specifically, to enhance the expressiveness of the point-to-shape auto-encoder, we propose a coarse-to-fine optimization strategy that initially reconstructs the coarse shape using only occupancy supervision and then refines by incorporating additional render-based perceptual loss.
Next, based on the geometrical covariant property of the point-to-shape auto-encoder and the appearance-invariant nature of image-to-shape diffusion, we establish a single-image-to-shape diffusion model with geometric alignment and generative rendering augmentation to enhance controllability and generalization ability.
For texture generation, we propose a geometry-conditioned ControlNet with reference attention fine-tuning that first generates multi-views consistent with the input image in both appearance and lighting. The corresponding PBR channels are estimated using our proposed multi-view PBR decomposer, then back-projected to UV space and processed through a UV-space inpainter to fill the invisible faces.
As a result of these advancements, \approach can generate PBR-textured 3D meshes with consistent geometry and exceptional fidelity within 30 seconds.
In summary, we introduce MeshGen, a novel pipeline for generating 3D meshes with geometric details and high-quality PBR textures. Key innovations of MeshGen include:
\begin{itemize}
    \item \textbf{Render-enhanced auto-encoder}. Our auto-encoder leverages a coarse-to-fine optimization with perceptual loss to incorporate surface information during training, which, combined with the designed ray-based regularization, ensures stability and substantially improves the quality of geometric representation.
    \item \textbf{Generative data augmentation.} Based on the analysis of geometry-texture disentangled frameworks, we introduce geometric alignment and generative rendering augmentation to image-to-shape diffusion training. The resulting model demonstrates strong controllability and generalization ability even with limited public datasets.
    \item\textbf{Consistent PBR-texture generator.} We design a texture generator that integrates multi-view geometric information while maintaining the visual fidelity of the reference image. When combined with our PBR decomposer, it efficiently produces highly consistent PBR materials.
\end{itemize}

\section{Related Work}
\label{gen_inst}
\subsection{3D Generation}
Early efforts in 3D generation focus on per-scene optimization methods based on CLIP similarity~\citep{CLIP, sanghi2021clip, jain2021dreamfields} and score distillation sampling~\citep{dreamfusion}. By utilizing powerful pre-trained image diffusion models, these methods soon excel in various 3D generation tasks~\citep{prolificdreamer, fantasia3d, magic3d, tang2023dreamgaussian, chen2024gsgen, shi2023MVDream, sweetdreamer, wang2023imagedream, sun2023dreamcraft3d, GaussianEditor}.
Despite great success has been achieved, optimization-based methods still suffer from slow generation speed and low success rates. 
To overcome these challenges, researchers have explored multi-view generation~\citep{liu2023one2345, tang2024mvdiffusionpp,Lu2023Direct25DT,liu2023syncdreamer,long2023wonder3d,wu2024unique3d,li2024era3d,chen2024v3d,voleti2024sv3d} and large reconstruction models~\citep{szymanowicz23splatter, liu2023unidream, DMV3D, xu2024agg, LRM, instant3d, PF-LRM, tang2024lgm, wang2024crm, TripoSR2024, sf3d2024, wei2024meshlrm}.
InstantMesh~\citep{xu2024instantmesh} adopts a two-stage optimization strategy that firstly trains a multi-view to triplane NeRF model, then uses this model as initialization for FlexiCubes~\citep{shen2023flexicubes}, thus yielding direct textured mesh reconstruction from images.
MeshFormer~\citep{liu2024meshformer} utilizes a hierarchical voxel structure for efficient large reconstruction model training.
Although these methods have advanced 3D generation in speed and quality,
the unsatisfying performance of multi-view generation and the growing demands for higher mesh quality have led researchers to focus increasingly on the development of native 3D generation methods~\citep{liu2023one2345++, gupta20233dgen, chen2024primx, Wang2023rodin, ren2024xcube,hong20243dtopia}.
3DShape2Vecset~\citep{3dshape2vecset} and CLAY~\citep{zhang2024clay} exploit latent vector sets as representation, significantly enhancing the expressiveness of the latent space.
CraftsMan~\citep{li2024craftsman} improves 3DShape2Vecset by incorporating point normal as input to the auto-encoder and proposes a normal enhancement process for finer details.
Direct3D~\citep{wu2024direct3d} exploits triplane as representation for capturing the structural 3D information.

\subsection{Texture Generation}
Initial efforts in mesh texturing have focused on only utilizing image diffusion models through iterative inpainting and optimization~\citep{Chen_2023_ICCV, jiang2024flexitex,cao2023texfusion}. TEXTure~\citep{TEXTure} presents an iterative texturing method that employs a pre-trained depth-to-image diffusion model to progressively refine a 3D model's texture map from various views. 
FlashTex~\citep{deng2024flashtex} proposes a light ControlNet for text-to-PBR generation. 
In addition to methods that use only image diffusion, various learning-based strategies initiate the training of generative texturing models using 3D textured mesh data~\citep{nichol2022point,luo2023scalable,jun2023shap,li20223d,collins2022abo,deitke2023objaverse,chen2022auv,yu2021learning,cheng2023tuvf,yu2023texture}.
Texturify~\citep{siddiqui2022texturify} proposes a GAN-based pipeline with face convolution to colorize meshes without direct supervision.
Paint3D~\citep{zeng2023paint3d} proposes a coarse-to-fine strategy that firstly colorizes sparse views with depth-based inpainting and then improves texture quality within the UV space.
Meta 3D TextureGen~\citep{bensadoun2024meta} exploits a geometry-conditioned multi-view generator for text-to-texture generation.

\begin{figure*}[!t]
    \vspace{-8mm}
    \centering
    \includegraphics[width=1.0\linewidth]{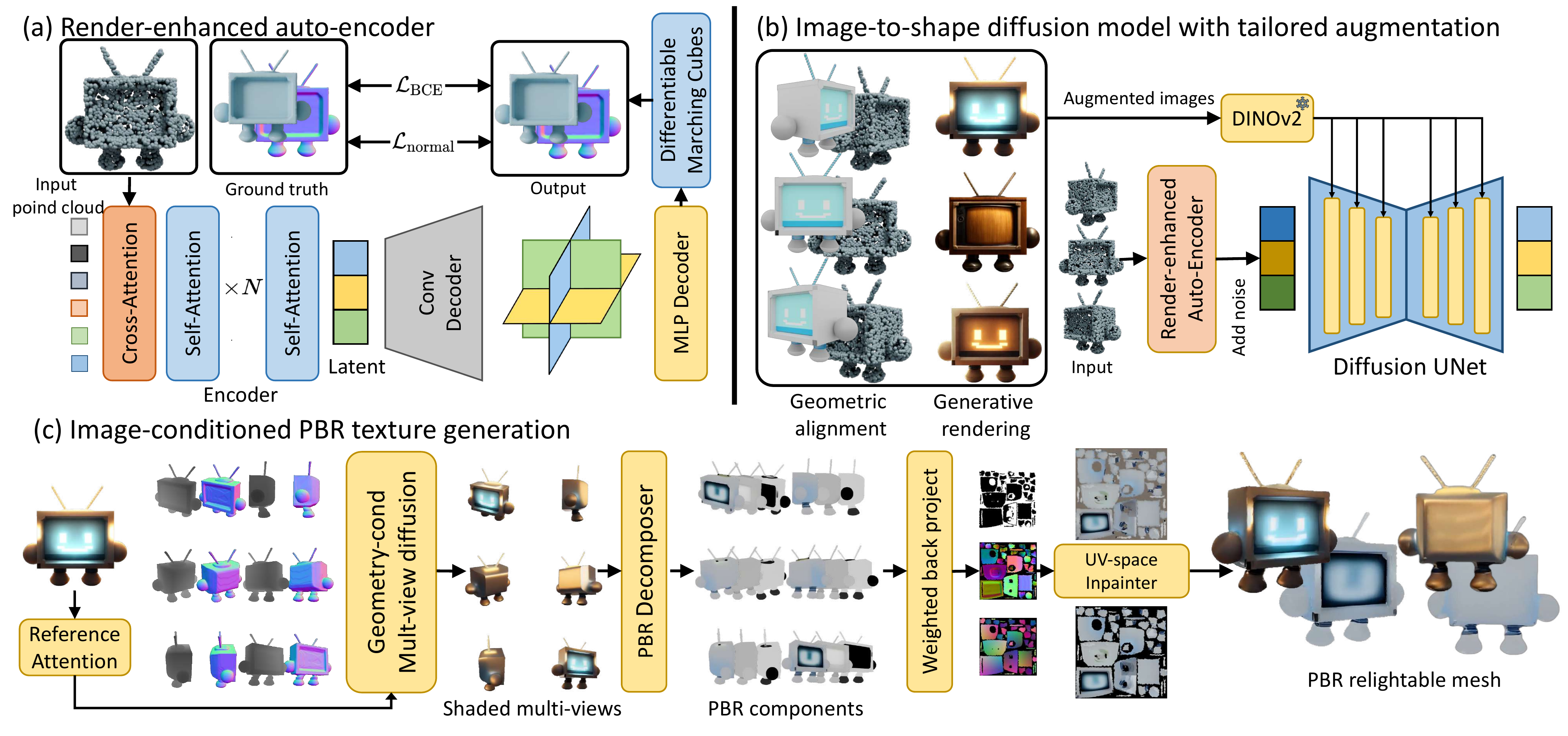}
    \vspace{-7mm}
    \caption{Overview of \approach. We first train a render-enhanced auto-encoder to compress meshes to more compact latent space (\sref{sec:autoencoder}). We establish an image-to-shape diffusion model based on our tailored generative augmentations for improving controllability and generalization ability (\sref{sec:diffusion}). The obtained mesh undergoes a reference attention-based multi-view synthesis and a PBR decomposer to obtain multi-view PBR channels. A UV-space inpainter is then exploited to fill the areas invisible in multi-view images (\sref{sec:texture}).}
    \label{fig:overview}
    \vspace{-3mm}
\end{figure*}

\section{Method}
The overall pipeline of \approach is demonstrated in \fref{fig:overview}. We first train a render-enhanced auto-encoder to compress the 3D meshes into compact triplanes. A single-image-to-shape diffusion model is then established based on the proposed geometric alignment and generative rendering augmentation to enhance controllability and generalization ability. The decoded 3D mesh then undergoes our texturing pipeline for PBR material generation. The detailed \approach methodology is presented as follows.

\subsection{Render-enhanced auto-encoder}
\label{sec:autoencoder}
\subtitle{Transformer-based point-to-shape auto-encoder.}
To compress the discrete triangle meshes into a continuous latent space, we adopted the same encoder as used in prior native 3D generation approaches~\citep{3dshape2vecset, zhang2024clay, li2024craftsman}, namely the point-to-shape encoder. 
For a given 3D object, we uniformly sample $N_P$ points from its surface and encode them using Fourier positional encoding~\citep{rahaman2019spectral}. We then introduce learnable queries to extract information via cross-attention, followed by self-attentions to refine the representation.
The complete encoding process can be formulated as
\begin{equation}
    \mathbf{z} = \texttt{SelfAttn}^{n}(\texttt{CrossAttn}(Q, \texttt{PE}(P))),
\end{equation}
where $n$ refers to the number of self-attention layers, \texttt{SelfAttn}, \texttt{CrossAttn} and \texttt{PE} represents self-, cross-attention, and Fourier positional encoding. Here $Q \in \R^{N_z \times d_z}$ and $P \in \R^{N_P \times 3}$ represent the learnable query set and the sampled point cloud respectively, $N_z$ and $d_z$ refer to the number of learnable queries and the dimension of the latent space.
To incorporate render-based perceptual loss during auto-encoder training, we opt for triplane as the latent representation~\citep{wu2024direct3d} over the latent vector set from 3DShape2Vecset~\citep{3dshape2vecset}, as it facilitates higher-resolution surface extraction by requiring only an MLP decoder for querying occupancy, rather than cross-attention with all latents.
To obtain the occupancy of a specific point, a convolutional decoder is applied to upsample the encoded latent to a higher resolution to represent finer details. As analyzed in \citep{Wang2023rodin, wu2024direct3d}, we concatenate the three planes in the height dimension instead of the channel dimension to avoid artifacts caused by spatial misalignment. The occupancy of point $\mathbf{x}$ can be formulated as
\begin{equation}
    \text{Occupancy}(\mathbf{x}) = \text{MLP}(\texttt{UpSample}(\mathbf{z}_{\text{tile}}), \mathbf{x}),
\end{equation}
where \texttt{Upsample} denotes the convolution-based upsampling network, MLP refers to the occupancy decoder, $\mathbf{z}_{\text{tile}}$ represents the height-concatenated triplane. 
\begin{figure*}[t]
    \vspace{-8mm}
    \centering
    \includegraphics[width=0.95\textwidth]{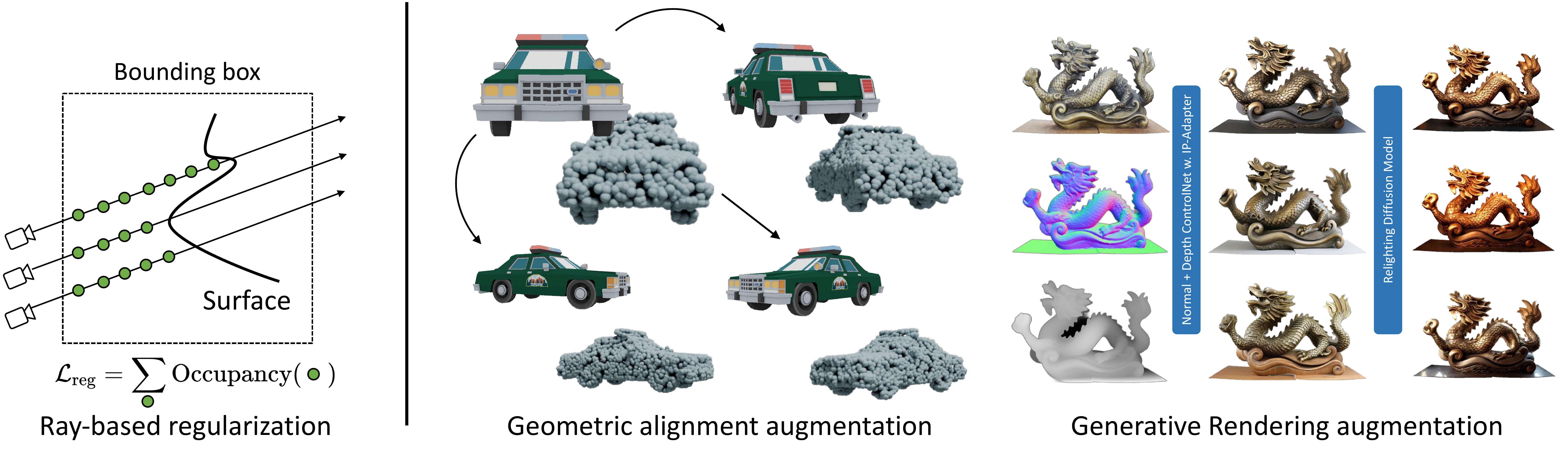}
    \vspace{-5mm}
    \caption{Illustration of the proposed ray-based regularization and two data augmentations.}
    \label{fig:augmentation}
    \vspace{-5mm}
\end{figure*}
\subtitle{Perceptual loss with ray-based regularization and coarse-to-fine optimization.}
Previous point-to-shape auto-encoders rely solely on occupancy, which results in poor performance when reconstructing high-frequency details. To address this, we propose enhancing the auto-encoder using normal maps with perceptual loss as additional supervision. During training, we query the occupancy of a \(256^3\) grid and extract the iso-surface differentiably~\citep{wei2023neumanifold}. 
However, early experiments show that simply incorporating render loss will cause severe floaters in the final output mesh (see \aref{app:more_ablations}), a problem also noted in previous research~\citep{wei2024meshlrm}. To mitigate this, we propose a ray-based regularization that forces the occupancy in empty spaces to approach zero. As illustrated in the left part of \fref{fig:augmentation}, for each camera ray, we uniformly sample \(N_s\) points between the ray-bounding box intersection and the surface point, enforcing their occupancy to be near zero. 
Due to the locality of differentiable marching cubes, the gradients of the render loss can only propagate to points near surface vertices. Therefore, a coarse-to-fine training process is necessary to ensure the effectiveness of the render loss. During the coarse stage, we apply standard binary cross-entropy (BCE) loss for the point-to-shape auto-encoder, along with a KL loss to regularize the latent space and a total variation loss~\citep{yu2021plenoxels} to reduce floaters, i.e.

\begin{equation}
    \mathcal{L}_{\text{coarse}}=\mathcal{L}_{\text{BCE}} + \lambda_{\text{KL}}\mathcal{L}_{\text{KL}} + \lambda_{\text{TV}}\mathcal{L}_{\text{TV}},
\end{equation}
where $\mathcal{L}_{\text{BCE}}$, $\mathcal{L}_{\text{KL}}$ and $\mathcal{L}_{\text{TV}}$ denote the BCE loss, the KL loss and the total variation loss respectively, $\lambda_{\text{KL}}$ and $\lambda_{\text{TV}}$ refer to the loss weights.
After the coarse stage training, the model is capable of reconstructing a coarse mesh from the input point cloud. In the refinement stage, we exploit render loss with ray-based regularization to enhance the details of the reconstructed mesh:
\begin{equation}
    \begin{aligned}
        \mathcal{L}_{\text{refine}}=&\mathcal{L}_{\text{BCE}} + \lambda_{\text{KL}}\mathcal{L}_{\text{KL}} + \lambda_{\text{TV}}\mathcal{L}_{\text{TV}} + 
        \lambda_{\text{MSE}}\mathcal{L}_{\text{normal}}^{\text{MSE}}\\
        &+\lambda_{\text{LPIPS}}\mathcal{L}_{\text{normal}}^{\text{LPIPS}} +
        \lambda_{\text{reg}}\mathcal{L}_{\text{reg}},
    \end{aligned}
\end{equation}
where $\mathcal{L}_{\text{normal}}^{\text{MSE}}$ and $\mathcal{L}_{\text{normal}}^{\text{LPIPS}}$ denotes the MSE and LPIPS~\citep{LPIPS} loss for rendered normal, $\lambda_{\text{MSE}}$, $\lambda_{\text{LPIPS}}$, and $\lambda_{\text{reg}}$ refers to the corresponding loss weight. The concrete hyper-parameter settings are presented in \aref{app:details}.

\subsection{Image-to-shape diffusion model with generative data augmentation}
\label{sec:diffusion}
\begin{figure*}[t]
    \vspace{-8mm}
    \centering
    \includegraphics[width=1.0\textwidth]{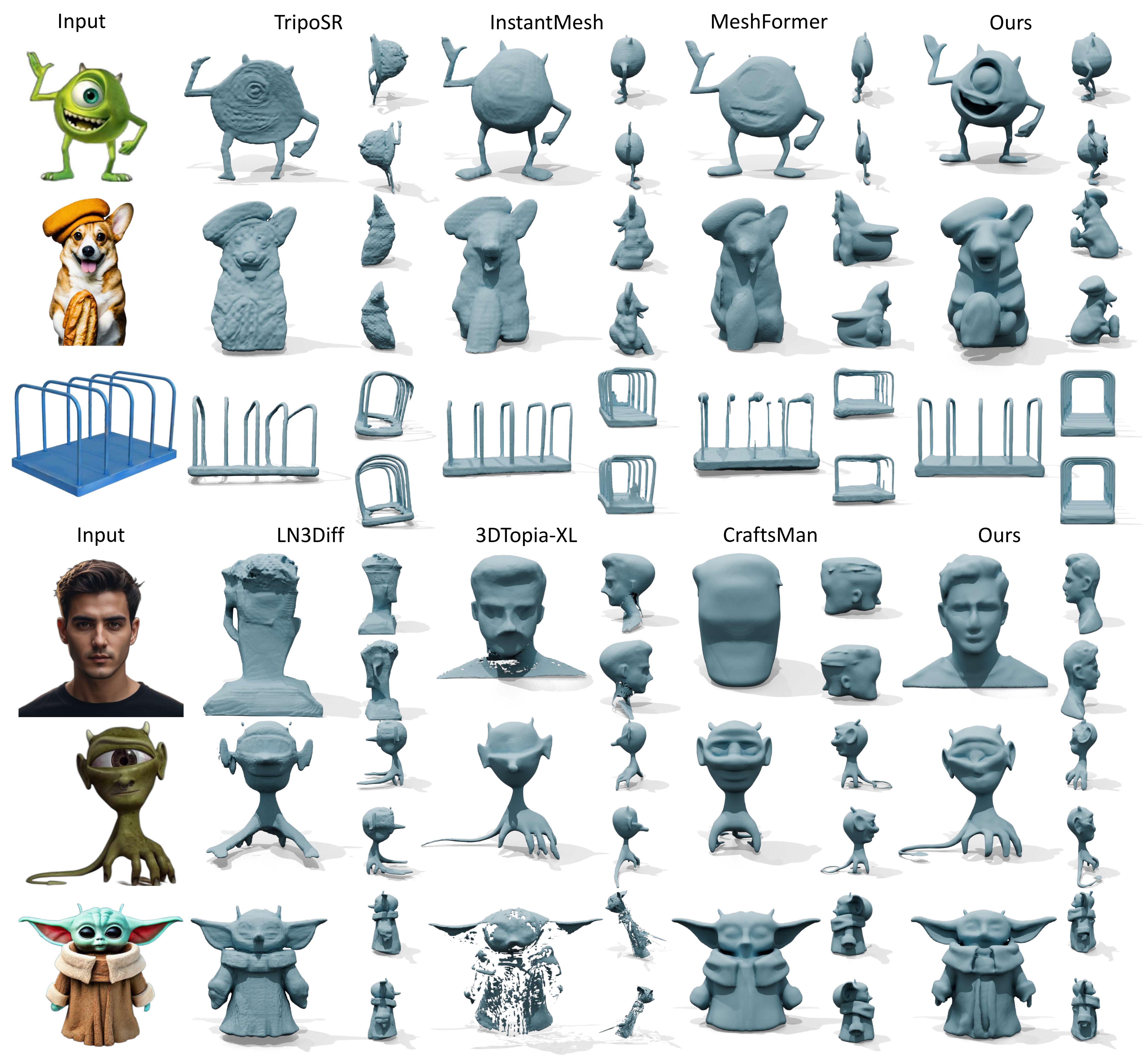}
    \vspace{-10mm}
    \caption{Qualitative comparison on in-the-wild images with state-of-the-art large reconstruction models (upper part, including InstantMesh~\citep{xu2024instantmesh}, TripoSR~\citep{TripoSR2024} and MeshFormer~\citep{liu2024meshformer}) and 3D native diffusion models (lower part, including CraftsMan~\citep{li2024craftsman}, LN3Diff~\citep{lan2024ln3diff}, and 3DTopia-XL~\citep{chen2024primx}). More comparisons and results are presented in the \aref{app:more_results} and supplement video.}
    \label{fig:comp_lrm}
    \vspace{-3mm}
\end{figure*}

As shown in \fref{fig:comp_lrm}, compared to large reconstruction models, existing native 3D generation models tend to generate simple and symmetrical shapes that differ significantly from the reference images.
We believe this phenomenon arises because the training of existing models predominantly relies on the Objaverse dataset~\citep{objaverse}, and a considerable portion of the objects in Objaverse have symmetrical geometry and lack realistic textures. Therefore, the diffusion models trained on it tend to replicate simple geometries and struggle to generalize to images with complex textures or lighting.
To obtain a diffusion model with strong generalization capabilities on limited public data, we identify two key differences between the point-to-shape auto-encoder based and the previous NeRF-based native 3D generation pipeline (typical methods include Rodin~\citep{Wang2023rodin}, 3DTopia~\citep{hong20243dtopia}, etc):
\subtitle{(1) Geometrical covariant auto-encoder.} 
Previous NeRF-based native 3D generation methods utilize a per-object optimized NeRF as the latent representation, which requires pre-computing and storing numerous latents and lacks geometric covariance, as it necessitates re-optimizing the radiance field to obtain triplane latent variables after applying a transformation to the object. In contrast, the point-to-shape auto-encoder takes point clouds as input and does not require per-object optimization, it naturally achieves geometric covariance for transformations such as rotations by directly manipulating the point cloud.
\subtitle{(2) Appearance invariant image-to-shape modeling.} 
Previous NeRF-based methods generate textured assets, causing entanglement between input images and textures. In contrast, our image-to-shape diffusion model maps images solely to shapes, meaning renderings under reasonable texture or lighting conditions should correspond to the same shape.

Based on both insights, we propose two data augmentations that are critical for training the image-to-shape model.

\subtitle{Geometric alignment augmentation.}
To enhance image-shape correspondence during training, we propose utilizing the geometric covariance property of our point-to-shape auto-encoder to ensure that different views of the same object correspond to different latents.
Specifically, for each object in the dataset, we select one view from multi-view images as the condition and rotate the point cloud's azimuth to align the object's orientation with the selected image as the target (see the middle part of~\fref{fig:augmentation} for a simple demonstration).
The aligned image-shape pairs are then used as training data for the diffusion model.
Our experiments reveal that geometric alignment not only expands the training dataset but also significantly improves the alignment between generated shapes and images. We present the corresponding ablation study in ~\fref{fig:ablation}.

\subtitle{Generative rendering augmentation.}
To enhance the generalization ability of the image-to-shape diffusion, we propose leveraging the appearance-invariant property by utilizing generative rendering to synthesize images with realistic textures and rich lighting based on the geometry of the object.
Concretely, for each rendered image in the dataset, we utilize the corresponding normal map and depth map as control signals to synthesize realistic renderings with ControlNet~\citep{zhang2023adding}. 
We then use IC-light~\citep{iclight} to generate renderings under various lighting conditions. Experiments show that generative rendering augmentation is highly beneficial for helping diffusion models understand lighting effects and generalize to real-world images (see \fref{fig:ablation} for the corresponding ablation study).

\subtitle{Image-to-3D diffusion UNet.} 
Based on the proposed two augmentations, we train a single-image-to-shape diffusion UNet on a filtered subset of GObjaverse~\citep{qiu2024richdreamer}. 
Our diffusion UNet is similar to Stable Diffusion~\citep{stablediffusion}, consisting of interleaved convolution and attention layers.
Following Rodin~\citep{Wang2023rodin}, we concatenate triplanes along the height dimension as input to the UNet to avoid spatial mismatches. Interactions between different planes are handled via self-attention layers. 
The input reference image is encoded using pre-trained DINOv2~\citep{oquab2024dinov} and injected with cross-attention.
Following SD3~\citep{esser2024scaling}, we adopt rectified flow~\cite{liu2022flow} with lognorm timestep sampling as the training schedule. For more details on the training and inference of our diffusion UNet, please refer to \aref{app:details}.

\subsection{Texture Generation}
\label{sec:texture}
Our texturing pipeline starts by generating multi-view shaded images using a geometry-conditioned ControlNet with reference attention. The corresponding PBR components are estimated using a PBR decomposer, then back-projected to UV space and inpainted to fill invisible areas. The detailed pipeline is as follows.

\subtitle{Geometry-conditioned multi-view generation with reference attention.}
The main challenge in our texturing pipeline is the imperfect alignment between generated shapes and input images, preventing direct back-projection or texture optimization. Previous methods resolve by using IP-adapter~\citep{ye2023ip-adapter} to incorporate reference image information, but this often fails to accurately resemble the reference image.
\begin{figure}[tb]
    \centering
    \vspace{-8mm}
    \includegraphics[width=0.45\textwidth]{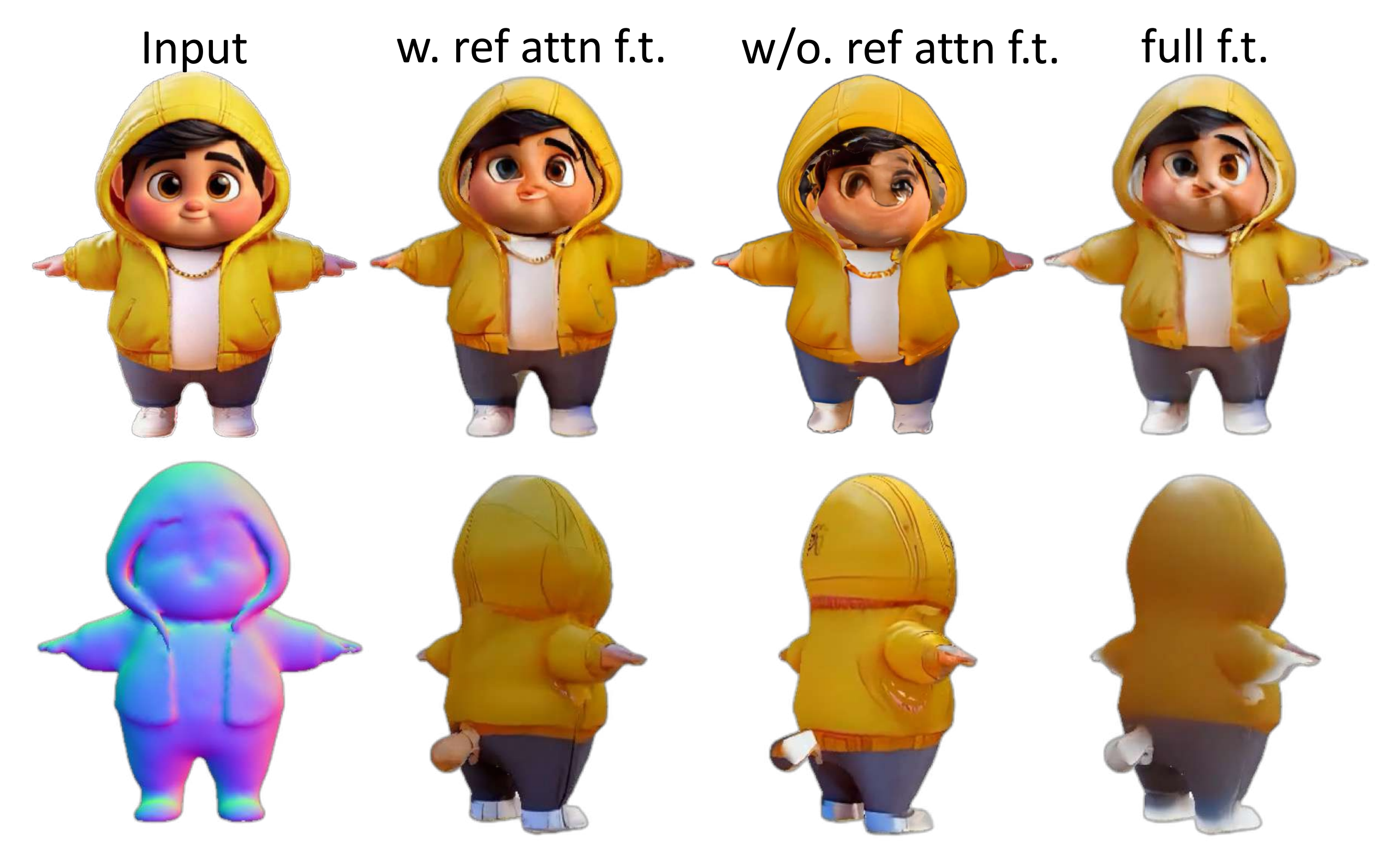}
    \vspace{-4mm}
    \caption{The effectiveness of the proposed reference attention fine-tuning. Here ``ref attn'' stands for reference attention, ``f.t.'' denotes fine-tuning.}
    \vspace{-5mm}
    \label{fig:ref_attn}
\end{figure}
In response, we propose using reference attention to inject information from the reference image.
Compared with IP-Adapter who exploit pre-trained CLIP features, reference attention~\citep{refattn} integrates the keys and values from self-attention layers corresponding to the reference image into the denoising process, thus enhancing the consistency between the generated and original images (see \aref{app:ref_attn} for detailed illustration).
To make the generation geometry-aware, we build a ControlNet~\citep{zhang2023adding} on top of the base model, enabling it to generate multi-view images according to multi-view normal and depth maps. 
However, applying this method alone does not yield satisfactory results(see ``w/o. ref attn f.t.'' in \fref{fig:ref_attn}). This is because the model prioritizes geometric correspondence over semantic information since the reference image and the geometry are perfectly aligned during training.
To address this, we introduce a reference fine-tuning stage after ControlNet training to enhance focus on semantic information. We simulate imperfect matches by randomly translating and rotating condition images, then fine-tune only the projection matrices of the reference attention's keys and values on the augmented dataset. As shown in \fref{fig:ref_attn}, this lightweight fine-tuning compensates for performance loss from imperfect matches without affecting generative capability, unlike full fine-tuning, which results in overly smooth textures, or the original model, which struggles with inconsistency.

\subtitle{Multi-view PBR decomposition.}
In order to generate relightable PBR textures, we propose a diffusion-based multi-view PBR decomposer, aiming to decompose the shaded multi-view image to PBR components with multi-view information. 
Specifically, inspired by \citep{rgbx}, our PBR decomposer employs an InstructPix2Pix-based~\citep{brooks2022instructpix2pix} architecture, concatenating the shaded image latent and the noisy latent along the channel dimension to output the desired PBR components, i.e.
\begin{equation}
    \mathbf{I}^{MV}_{i}(y)=g_{\phi}(\mathbf{I}^{MV}_{i-1}(y);\mathbf{I}^{MV},\mathbf{I}^{\text{ref}},i,\tau(y)),
\end{equation}
where $g_{\phi}$ represents the denoising UNet for PBR decomposer, $\tau$ denotes the pre-trained CLIP~\citep{CLIP} text encoder, $y \in \{\texttt{"metallic"}, \texttt{"roughness"}, \texttt{"albedo"}\}$ denotes the component prompt for PBR texture, $\mathbf{I}^{MV}_{i}(y)$ refers to the denoised $y$ component at denoise step $i$, $\mathbf{I}^{MV}$ and $\mathbf{I}^{\text{ref}}$ denote the multi-view shaded image and the reference image respectively.
After generating multi-view PBR components, we use a view-weighted approach to fuse the multi-view textures in UV space, i.e. $\text{UV}=\sum_i\text{Softmax}_i(\text{BP}(\mathbf{I}^{(i)}), \text{BP}(w^{(i)}))$
, where $\text{BP}$ refers to back-projecting the rendered image to UV space, $\mathbf{I}^{(i)}$ denote the target image for the $i$-th view and $w^{(i)}$ represents the pixel-wise weight calculated as the cosine of the viewing angle to the point.

\begin{figure*}[t]
    \vspace{-7mm}
    \centering
    \includegraphics[width=1.0\textwidth]{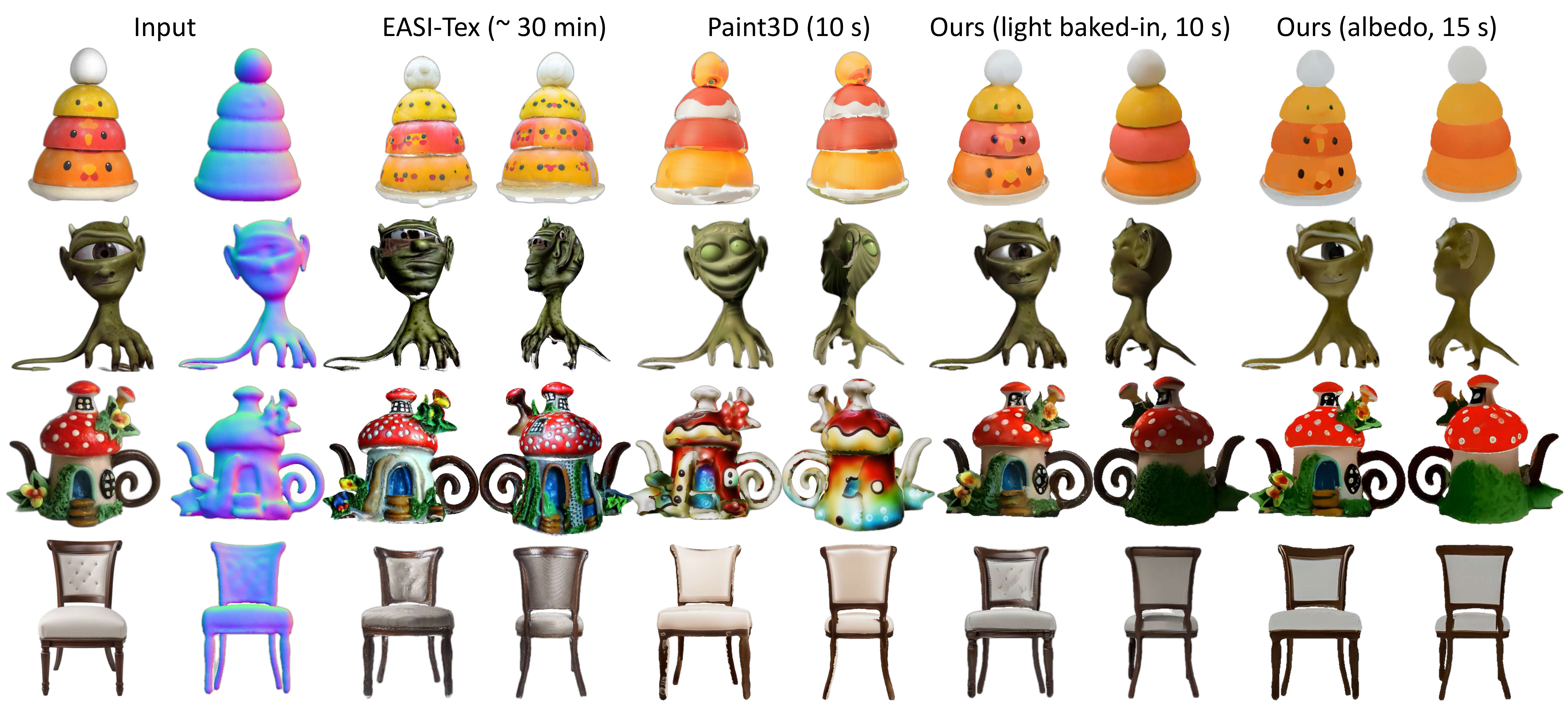}
    \vspace{-3mm}
    \caption{Qualitative comparison with previous mesh texturing methods, including EASI-Tex~\citep{perla2024easitex} and Paint3D~\citep{zeng2023paint3d}.}
    \label{fig:comp_texture}
    \vspace{-4mm}
\end{figure*}

\subtitle{UV-space texture inpainting.}
For meshes with complicated topology, the generated views are not adequate to cover the entire surface of the mesh. We propose a UV-space texture inpainter to fill the invisible part of the multi-views. Specifically, due to the significant gap between casual images and texture maps, we first train a LoRA on the texture maps in Objaverse.
Subsequently, we merge the LoRA into the original UNet and train an inpainting ControlNet on top of it. To let the texture map perceive spatial information, our control signal includes not only the masked image but also the normal and position maps in UV space. 
We present more details about the texturing pipeline in \aref{app:pbr_details}.

\begin{table}[b]
\centering
\vspace{-3mm}
\caption{Quantitative comparison with previous methods on GSO~\citep{GSO} and OmniObject3D~\citep{wu2023omniobject3d}. Here CD denotes Chamfer Distance and FS stands for F-Score. The best result is highlighted in \textbf{bold}, while the second-best result is \underline{underscored}. As \approach is a generative model, we report the mean and standard deviation across five different random seeds in the ``Ours'' row.}
\label{tab:comp}
\vspace{-3mm}
\resizebox{0.48\textwidth}{!}{%
\begin{tabular}{l|cccccccccccccccc}
\toprule[2pt]
\multirow{2}{*}{Method} & \multicolumn{2}{c}{GSO} & \multicolumn{2}{c}{OmniObject3D} \\
    & FS $\uparrow$ & CD $\downarrow$ & FS $\uparrow$ & CD $\downarrow$\\
    \midrule
    One2345++~\citep{liu2023one2345++} & 0.936 & 0.039 & 0.871 & 0.054  \\ 
    TripoSR~\citep{TripoSR2024} & 0.896 & 0.047 & 0.895 & 0.048  \\ 
    CRM~\citep{wang2024crm} & 0.886 & 0.051 & 0.821 & 0.065  \\ 
    LGM~\citep{tang2024lgm} & 0.776 & 0.074 & 0.635 & 0.114  \\ 
    InstantMesh~\citep{xu2024instantmesh} & 0.934 & 0.037 & 0.889 & 0.049  \\ 
    MeshLRM~\citep{wei2024meshlrm} & 0.956 & 0.033 & 0.91 & 0.045  \\ 
    MeshFormer~\citep{liu2024meshformer} & \underline{0.963} & \underline{0.031} & \underline{0.914} & \underline{0.043} \\ 
    CraftsMan~\citep{li2024craftsman} & 0.702 & 0.088 & 0.599 & 0.143 \\ 
    LN3Diff~\citep{lan2024ln3diff} & 0.647 & 0.102 & 0.534 & 0.168 \\ 
    \midrule
    Ours & {\footnotesize\textbf{0.971}}{\scriptsize $\pm$.014} & {\footnotesize\textbf{0.028}}{\scriptsize $\pm$.005} & {\footnotesize\textbf{0.918}}{\scriptsize $\pm$.010} & {\footnotesize\textbf{0.040}}{\scriptsize $\pm$.004} \\
\bottomrule[2pt]
\end{tabular}
}
\end{table}

\section{Experiments}
\subsection{Mesh Generation}

In our experiments, we compare our method with state-of-the-art image-to-3D methods from two categories. (1) \textbf{Large reconstruction models}~\citep{LRM, PF-LRM, instant3d, wang2024crm, tang2024lgm} exploit a neural network to map sparse views into 3D representations, including InstantMesh~\citep{xu2024instantmesh}, TripoSR~\citep{TripoSR2024} and MeshFormer~\citep{liu2024meshformer}. To be clear, the results of MeshFormer are obtained through their official demo, since has not been open-sourced.
(2) \textbf{Native 3D generation models}~\citep{liu2023one2345++, 3dshape2vecset, zhang2024clay} compress discrete 3D meshes into a continuous and compact latent space using a 3D auto-encoder, followed by a diffusion model trained on this latent space to achieve 3D generation, including LN3Diff~\citep{lan2024ln3diff}, 3DTopia-XL~\citep{chen2024primx}, and CraftsMan~\citep{li2024craftsman}.

\subtitle{Qualitative results.}
We present a qualitative comparison between our method and large reconstruction models in the upper part of \fref{fig:comp_lrm}. Our approach significantly outperforms others in generating high-quality geometry. Specifically, our method excels at producing complex structures from images, such as the mouth of the alien and the file sorter, where large reconstruction models struggle. Additionally, large reconstruction models often suffer from poor multi-view generation outcomes, resulting in lower quality when reconstructing details that require multi-view consistency.
In the lower part of \fref{fig:comp_lrm}, we provide a qualitative comparison with 3D native generation methods. As we mentioned earlier, previous native 3D generation models are limited by data and the expressive power of the autoencoder, often resulting in overly simple and symmetrical geometries that significantly differ from the reference images. 
Our method, leveraging the proposed two data augmentations, effectively addresses this issue. This image-shape alignment enhances the controllability of shape generation and simplifies subsequent texturing. 

\subtitle{Quantitative results.} In \tref{tab:comp}, we demonstrate quantitative comparisons with previous methods. For the evaluation setting, we follow MeshFormer~\citep{liu2024meshformer}, calculate F-Score (with a threshold of 0.2) and Chamfer distance on GSO~\citep{GSO} and OmniObject3D~\citep{wu2023omniobject3d}. 
As a generative model, our results include a certain degree of diversity, so we report the mean and variance of the model's performance across five random seeds. \approach outperforms others in both metrics, demonstrating the effectiveness of our method.

\begin{figure*}[t]
    \centering
    \vspace{-4mm}
    \includegraphics[width=0.9\linewidth]{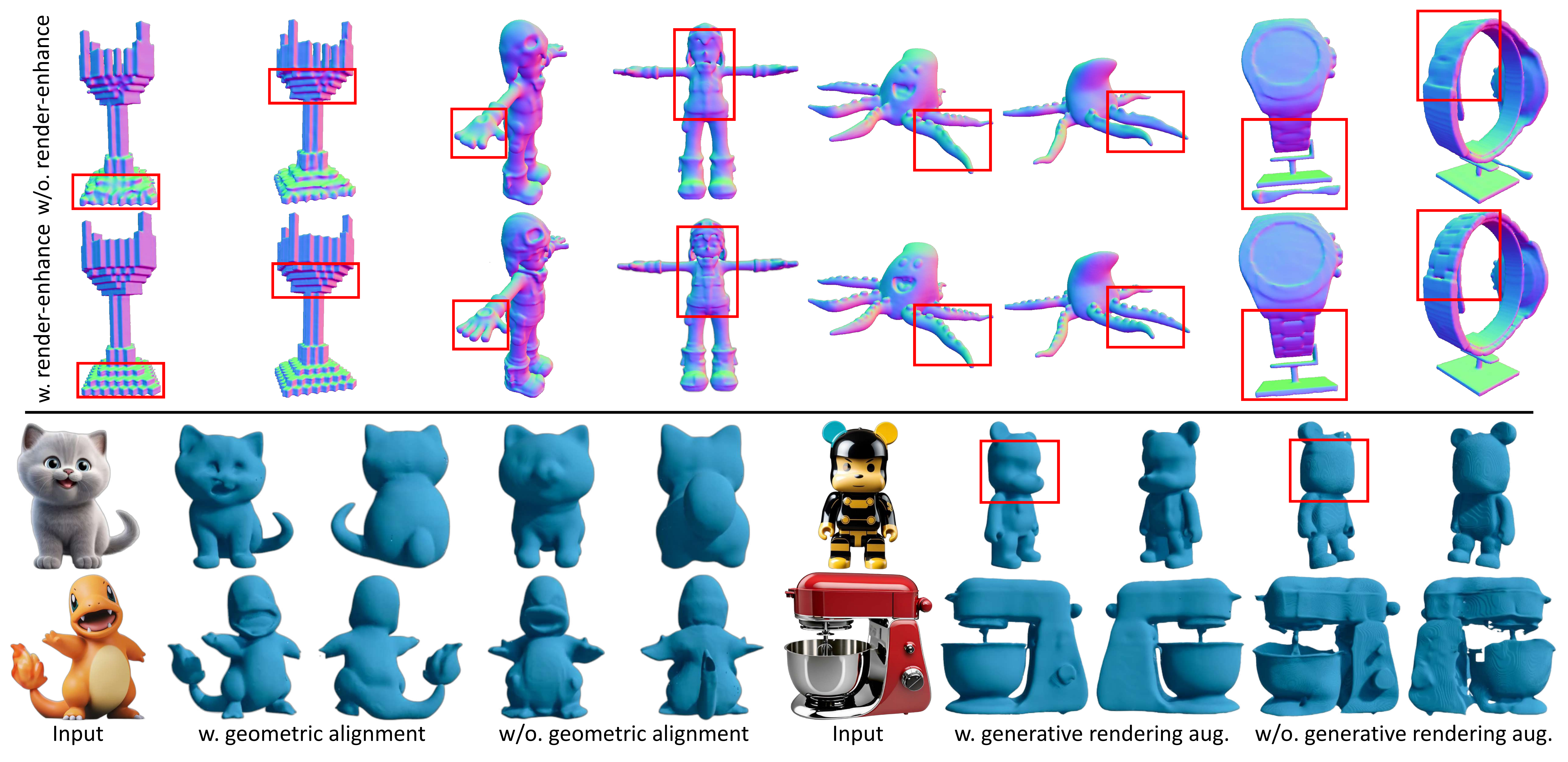}
    \vspace{-5mm}
    \caption{Ablation study results on render-enhanced auto-encoder, geometric alignment, and generative rendering augmentation.}
    \label{fig:ablation}
    \vspace{-4mm}
\end{figure*}

\subsection{Texture Generation}

\begin{table}[b]
    \vspace{-4mm}
    \centering
    \caption{User study results on image-conditioned texture generation, the value refers to the win rate in percentage.}
    \label{tab:texture_user_study}
    \vspace{-2mm}
    \resizebox{0.48\textwidth}{!}{%
    \begin{tabular}{c|ccccccccccccccccccccc}
        \toprule[2pt]
        Metric & TEXTure & EASI-Tex & Paint3D & Ours \\
        \midrule[1pt]
         Image Alignment (\%) & 1.92 & 3.85 & 1.92 & \textbf{92.31} \\
         Overall Quality (\%) & 3.85 & 7.69 & 5.77 & \textbf{82.69} \\
        \bottomrule[2pt]
    \end{tabular}
    }
\end{table}

We first qualitatively compare our method with state-of-the-art image-conditioned mesh texturing approaches, including EASI-Tex~\citep{perla2024easitex} and Paint3D~\citep{zeng2023paint3d}.
\fref{fig:comp_texture} demonstrates the texturing results on meshes generated by our image-to-shape diffusion model. Our method significantly outperforms previous approaches in quality and texture-image consistency, even when the shape and image do not perfectly align.
In contrast, despite using image inversion which requires additional per-input optimization, EASI-Tex still struggles to maintain consistency with the original image and takes dozens of times longer than our approach.
Paint3D, which uses simple back projection and inpainting, exhibits noticeable seams in the generated textures and is prone to the Janus problem.
To quantitatively compare \approach texturing pipeline with previous methods, we conduct a user study to assess the performance with human ratings. Specifically, we ask 52 volunteers to choose the best method from TEXTure, EASI-Tex, Paint3D and ours in (a) Image alignment: how well does the texture align with the reference image; (b) Overal quality: the clarity, detail, and aesthetic appeal of the generated texture. \tref{tab:texture_user_study} demonstrates the win rate of each method in the two criteria. \approach is regarded as the most convincing model of all and significantly outperforms previous methods in both alignment and fidelity.
\begin{table}[b]
    \vspace{-3mm}
    \centering
    \caption{Quantitative ablation study on 3D auto-encoder. Here Acc, VIoU refer to accuracy and VolumeIoU on Objaverse test set. ``Ours(w/o. refine)'' denotes the no-render-loss variant.}
    \label{tab:autoencoder}
    \vspace{-2mm}
    \begin{tabular}{c|ccccccccccccccccccccc}
        \toprule[2pt]
        Metric & Vecset~\citep{3dshape2vecset} & Ours(w/o. refine) & Ours\\
        \midrule[1pt]
         Acc (\%) & 94.844 & 94.745 & \textbf{96.987} \\
         VIoU (\%) & 87.426 & 87.164 & \textbf{91.045} \\
        \bottomrule[2pt]
    \end{tabular}
\end{table}

\subsection{Ablations}

\subtitle{Render-enhanced auto-encoder.}
To evaluate the significance of incorporating render loss in the point-to-shape auto-encoder, we trained a variant without render-based perceptual loss and compared it to our render-enhanced version, as illustrated in the upper part of \fref{fig:ablation}. The render-enhanced auto-encoder performs markedly better, particularly in capturing high-frequency details such as the suckers on tentacles and the gaps in watch bands. In \tref{tab:autoencoder}, we also demonstrate through quantitative results the superiority of the render-enhanced auto-encoder compared to 3DShape2VecSet and the no-render-loss variant.

\subtitle{Geomtric alignment augmentation.}
To validate the impact of geometric alignment augmentation, we trained a variant without this augmentation for 300 epochs as an ablation study. The comparison is presented in the lower left part of \fref{fig:ablation}. Evidently, the diffusion model trained without geometric alignment augmentation tends to generate symmetric objects, whereas our model produces shapes that align well with the input images, significantly enhancing the model's controllability.

\subtitle{Generative rendering augmentation.}
To assess the impact of generative rendering augmentation on image-to-shape diffusion training, we trained a smaller model without this augmentation, as depicted in the lower right part of \fref{fig:ablation}. The model trained without generative rendering augmentation exhibits poor performance in handling lighting effects in images and struggles to infer the geometric structure based on lighting cues, which suggests that generative rendering augmentation significantly enhances the model's ability to understand lighting effects and interpret real-world images.

More ablations regarding the auto-encoder, image-to-shape diffusion model, and texture generation model are presented in \aref{app:more_ablations}. 

\section{Conclusion and Limitation}
\subtitle{Conclusion.}
In this paper, we propose \approach, a novel pipeline for generating delicate PBR textured mesh given a single image. \approach encodes 3D meshes with a render-enhanced auto-encoder. Based on the analysis of auto-encoder and image-to-shape diffusion, we propose to train the diffusion model with geometric alignment and generative rendering augmentation to enhance controllability and generalization ability. To generate PBR texture consistent with the image, we establish a reference attention-based multi-view generator followed by a PBR decomposer to obtain PBR components and a UV-space inpainter to fill the invisible part. Extensive experiments have demonstrated the effectiveness of our method. 

\subtitle{Limitation.} \approach demonstrates several limitations regarding texture quality and handling transparent objects. Concrete failure cases are shown in \aref{app:limitation}.

\section{Acknowledgment}
The author would like to thank Min Zhao for her help in the rebuttal phase. This work was jointly supported by National Natural Science Fund for Distinguished Young Scholars under Grant 62025304, Youth Fund of the National Natural Science Foundation of China under Grant 62306163, and was also supported by the Seed Funding Project of the second phase of the Tsinghua University (Department of Computer Science and Technology)-Siemens Ltd., China Joint Research Center for Industrial Intelligence and Internet of Things (JCIIOT).

\clearpage
{
    \small
    \bibliographystyle{ieeenat_fullname}
    \bibliography{main}
}

\input{supplemental}

\end{document}

%% file: preamble.tex
\usepackage{amsmath,amsfonts,bm}

\def\figref#1{figure~\ref{#1}}

\def\eqref#1{equation~\ref{#1}}

\def\1{\bm{1}}

\DeclareMathAlphabet{\mathsfit}{\encodingdefault}{\sfdefault}{m}{sl}
\SetMathAlphabet{\mathsfit}{bold}{\encodingdefault}{\sfdefault}{bx}{n}

\newcommand{\R}{\mathbb{R}}



%% file: supplemental.tex
\clearpage
\setcounter{page}{1}
\maketitlesupplementary
\appendix

\section{Implementation details}
\label{app:details}
\subsection{Auto-encoder}
\subtitle{Hyper-parameters.}
We present our hyper-parameter setting in training auto-encoder in \tref{tab:hyper_params}.
\begin{table*}[bht]
\begin{center}
\caption{Concrete hyper-parameter setting of our render-enhanced auto-encoder.}
\label{tab:hyper_params}
    \begin{tabular}{llcccccccccccccc}
        \toprule[2pt]
        
        Symbol  & Meaning & Value \\
        \midrule
        $N_P$ & Number of points sampled from a mesh & 65536 \\ 
        $N_z$ & Number of learnable queries & 3072 \\
        $n$ & Number of self-attention layers & 10\\
        $d_z$ & Dimension of the latent space & 16 \\
        $N_s$ & Number of samples for calculating ray-based regularization loss & 128 \\
        $\lambda_{\text{KL}}$ & Loss weight for KL loss & $10^{-6}$ \\
        $\lambda_{\text{TV}}$ & Loss weight for TV loss & $5\times 10^{-3}$ \\
        $\lambda_{\text{MSE}}$ & Loss weight for normal MSE loss & $1.0$ \\
        $\lambda_{\text{LPIPS}}$ & Loss weight for normal LPIPS loss & $2.0$ \\
        $\lambda_{\text{reg}}$ & Loss weight for ray-based regularization loss & $0.5$ \\
        \bottomrule[2pt]
    \end{tabular}
    \end{center}
\end{table*}

\subtitle{Data preparation.}
To calculate occupancy, the target mesh needs to be watertight, but most meshes in Objaverse do not meet this requirement. Therefore, we need to convert non-watertight meshes into watertight ones. Specifically, for each non-watertight mesh, we follow CLAY~\citep{zhang2024clay}, first computing an unsigned distance field with a resolution of $512^3$. Then, we use the marching cubes algorithm with a threshold of $2/512$ to extract the isosurface. To avoid thin surfaces inside and thin surfaces, we mark all parts not connected to the outermost region as internal, which can be quickly achieved using a connected component labeling algorithm. With off-the-shelf CUDA-based tools like torchcumesh2sdf~\citep{torchcumesh2sdf} and cc\_torch~\citep{cc_torch}, the preprocessing of a mesh can be done within 0.1s on a single GPU.

\subtitle{Training data.}
We train our \approach auto-encoder on a filtered subset of Objaverse~\citep{objaverse} consisting of approximately 150k triangle meshes.
For each mesh, we first normalize into $[-1, 1]^3$ and uniformly sample 65536 surface points on the mesh surface as the input of our auto-encoder. We then sample 100,000 spatial points near the surface by adding Gaussian noise with 0.01 std to the surface points. For random spatial points, we use stratified sampling to sample in a $64^3$ grid uniformly. We pre-compute and store surface points, sampled spatial points, and corresponding binary occupancy values for training efficiency.

\subtitle{Training details.}
We first train our auto-encoder for 150 epochs in the coarse stage with a batch size of 192. 
During training, we select an equal number of near points and random points for supervision~\citep{3dshape2vecset}.
The model obtained after the coarse stage can reconstruct the rough shape of the original mesh but lacks details. We then train the auto-encoder for another 50 epochs with the proposed render loss and ray-based regularization incorporated and a batch size of 16. The whole training process lasts 6 days on 8 NVIDIA A100 GPUs.

\subsection{Image-to-shape diffusion model}
\subtitle{Generative rendering augmentation.}
In generative rendering data augmentation, to enhance the similarity between the generated images and the original image, in addition to using the normal depth ControlNet and IP-adapter, we set the initial noise to the latent of the original image with maximum noise added. For relighting diffusion, we used IC-light~\citep{iclight}. Specifically, during data augmentation, we randomly select one lighting direction from the pre-defined light initial latent in IC-light (i.e., uniformly select from left, right, top, and bottom), and choose one lighting condition from a set of predefined light prompts. For filtering out low-quality images, we trained an MLP evaluator on 500 samples using CLIP embeddings of the original and relighted images to estimate quality scores. This straightforward method achieves 91\% accuracy on the validation set (100 samples).

\subtitle{Model setup.}
Our diffusion UNet takes in the noised triplane latent and exploits 8 ResNet blocks with spatial self-attention as the encoder and a symmetric architecture as the decoder.
We exploit DINOv2-G~\citep{oquab2024dinov} to encode the input image and inject the extracted feature to the diffusion UNet using cross-attention~\citep{stablediffusion}.
For the diffusion schedule, we follow SD3~\citep{esser2024scaling} to use the simple yet effective rectified flow~\cite{liu2022flow} with timesteps sampled from a standard logit-normal distribution.

\subtitle{Training data.}
We train the image-to-shape diffusion model with our proposed augmentations on a filtered subset of GObjaverse~\citep{qiu2024richdreamer}, which consists of about 120k high-quality meshes. During training, we randomly select one view as condition for the diffusion UNet and apply our proposed augmentations for both geometric and image enhancement.

\subtitle{Training details.}
To handle input images with different elevations, since the meshes in Objaverse are aligned in the gravity axis, we force the diffusion to generate meshes with an absolute elevation equal to zero. We experimentally found that this conditioning method works better than generating meshes with a rotation in elevation, as suggested in ~\citet{chen2024v3d}. We train the diffusion UNet of 16 NVIDIA A800 GPUs using bf16 precision with an effective batch size of 1536. The whole training lasts for about 18 days.

\subsection{PBR texture generation}
\label{app:pbr_details}

\begin{figure}[t]
    \centering
    \includegraphics[width=1.0\linewidth]{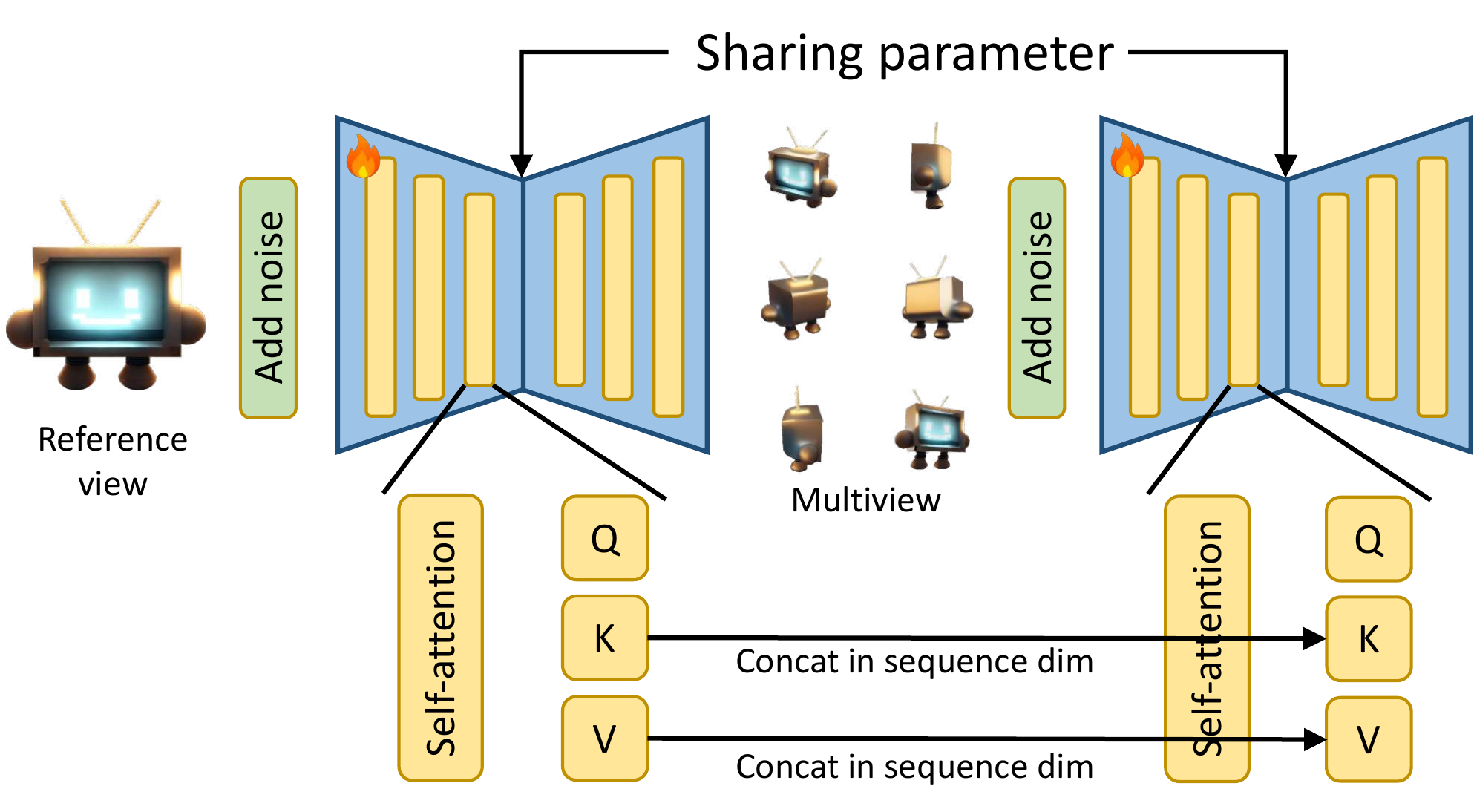}
    \caption{An illustration of reference attention.}
    \label{fig:ref_attn_demo}
\end{figure}

\subtitle{Reference attention.}
\label{app:ref_attn}
We demonstrate how reference attention works in \figref{fig:ref_attn_demo}. To condition the diffusion model on the reference image, we pass the noised reference image (at the same noise levels as the denoising latent) into the same diffusion UNet to obtain the key and value tensors of the reference image in self-attention layers. During sampling, the key and value tensors of the reference image are appended to the key and value tensors of the multi-views, enabling the diffusion model to perceive the reference image with more fine-grained information. This technique is also used in image editing~\citep{refattn} and video generation~\citep{hu2023animateanyone}.

\subtitle{Data preparation.}
To train the geometry-conditioned ControlNet and the multi-view PBR decomposer, we render multi-view images along with corresponding normals, depth, albedo, roughness, and metallic maps from a subset of Objaverse containing PBR materials using Blender. This creates a dataset of approximately 35k multi-view images. For UV space inpainting, we calculate multi-view visible masks and back-project them into UV space to identify the invisible parts of the texture map, rendering the UV space position and normal maps. To enhance robustness, we randomly erode the visible masks in both pixel and UV spaces.

\subtitle{Geometry-conditioned ControlNet.}
Our geometry-conditioned sparse view generator is built upon Zero123++~\citep{shi2023zero123plus}, which generates six views according to a given front view. Specifically, Zero123++ generates a 3x2 image grid by fine-tuning Stable Diffusion~\citep{stablediffusion} on Objaverse renderings.
To ensure the model perceives precise depth and positional information, we did not transform depth to normalized disparity as done in the original depth ControlNet~\citep{zhang2023adding}; instead, we performed a unified multi-view normalization based on camera distance and object bounding box. Specifically, the depth map is processed as $D_{\text{normalized}}=\frac{D-\text{bias}}{\text{scale}}$, where bias equals to camera distance minus the length of the diagonal of the bounding box (i.e. the minimal possible depth value) and the scale equals to the length of the diagonal of the bounding box.

\begin{figure*}
    \centering
    \includegraphics[width=0.9\linewidth]{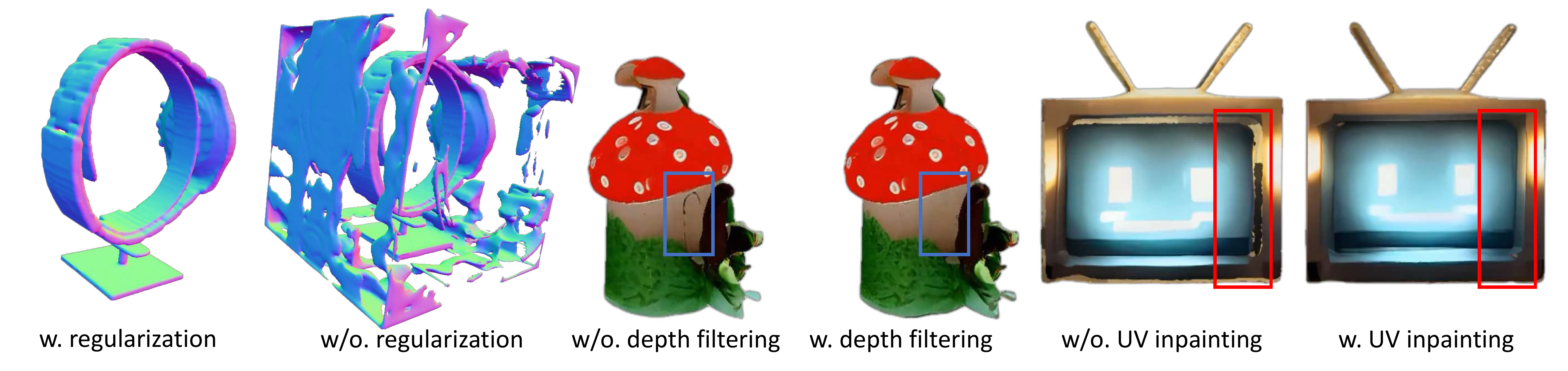}
    \vspace{-4mm}
    \caption{Ablations on ray-based regularization, depth filtering, and UV space inpainting.}
    \label{fig:more_ablations}
\end{figure*}

\subtitle{Multi-view target back-projection.}
As detailed in the main text, the obtained multi-view PBR components are merged in UV space using back-projection with softmax. We apply a softmax operation with a temperature of 0.1 to ensure consistent textures. However, images generated by ControlNet sometimes extend beyond object boundaries, causing some pixels to be back-projected onto surfaces behind them, leading to artifacts on the final texture map. To address this, we propose a simple depth filtering technique. For each view, we identify locations in the depth map where sudden changes occur and exclude these pixels during back-projection. Our experiments demonstrate that this approach effectively reduces artifacts, and the color values of the corresponding surface points can be supplemented by other views, as shown in the middle of \fref{fig:more_ablations}.

\subtitle{PBR decomposer.}
Our PBR decomposer is built on Zero123++ and incorporates multi-view shaded images using an InstructPix2Pix-based architecture. Specifically, the latent representation of the multi-view shaded image is concatenated along the channel dimension with the noisy latent. Each PBR channel is generated by the diffusion model using specific textual prompts. For instance, the model generates component \( y \) using the prompt \( y \), where \( y \in \{\texttt{"metallic"}, \texttt{"roughness"}, \texttt{"albedo"}\}. \)

\subtitle{UV space inpainting.}
Our UV space inpainter is a multi-channel ControlNet trained on top of the LoRA fine-tuned Stable Diffusion 1.5~\citep{stablediffusion}. The input to our inpainting model is a 9-channel image: the first three channels represent the normal map in UV space, the middle three channels represent the position map, and the last three channels contain the masked texture map, with pixel values set to -1 in regions that requires inpainting. During inference, we follow ControlNet inpainting~\citep{zhang2023adding}, applying masking in the latent space to maintain consistency in areas that do not require inpainting.

\section{More experiments}
\subsection{More ablations}
\label{app:more_ablations}
\begin{figure}[h]
    \centering
    \includegraphics[width=0.9\linewidth]{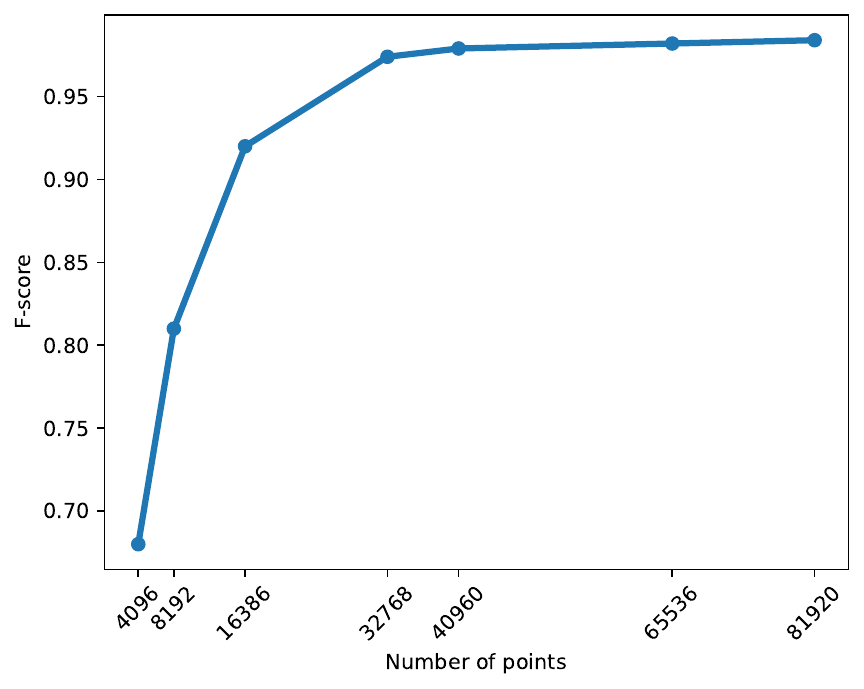}
    \vspace{-4mm}
    \caption{Ablations on the number of points used in mesh auto-encoding.}
    \label{fig:vae_perf_vs_points}
\end{figure}
\subtitle{Number of points in auto-encoder.}
We use 65536 points per mesh in auto-encoding. This decision stems from observing that VAE
reconstruction quality saturates at a high number of points (\fref{fig:vae_perf_vs_points}).

\subtitle{Quantitative ablation of generative data augmentation.}
\begin{table}[!t]
    \centering
    \caption{\textbf{Quantitative ablation study on the proposed data augmentation and comparison with Direct3D.} FS and CD denote f-score and chamfer distance. FS$_{\text{sym}}$, FS$_{\text{asym}}$, and FS$_{\text{complex}}$ represent f-scores for symmetric, asymmetric objects, and complex lighting images. Parentheses indicate metric changes for each modification.}
    \vspace{-3mm}
    \scalebox{0.73}{
    \begin{tabular}{c|cc|cc|c} 
    \toprule[2pt]
    Method & FS$\uparrow$ & CD$\downarrow$ & FS$_{\text{sym}}\uparrow$ & FS$_{\text{asym}}\uparrow$ & FS$_{\text{complex}}\uparrow$ \\
    \midrule[1pt]
+ render-ehanced VAE   & 0.907 & 0.061 & 0.932 & 0.881\scriptsize{(+.04)} & 0.841\scriptsize{(+.008)} \\
+ generative rendering & 0.918 & 0.053 & 0.944 & 0.892\scriptsize{(+.05)} & \textbf{0.897}\scriptsize{\textbf{(+.064)}} \\
+ geometric alignment  & 0.955 & 0.035 & 0.962 & \textbf{0.947}\scriptsize{\textbf{(+.11)}} & 0.855\scriptsize{(+.020)} \\
\midrule[1pt]
Ours (full) & \textbf{0.970} & \textbf{0.028} & \textbf{0.974} & \textbf{0.965} & \textbf{0.902} \\
    \bottomrule[2pt]
    \end{tabular}}
    \label{tab:data_augmentation_ablation}
    \vspace{-4mm}
\end{table}
As shown in \tref{tab:data_augmentation_ablation}, geometric alignment enhances performance on asymmetric objects, and generative rendering improves results under images with complex lighting conditions. Compared to Direct3D, each model design leads to significant improvements across all metrics.

\subtitle{The effectiveness of ray-based regularization.}
We show in the left part of \fref{fig:more_ablations} an example obtained using an auto-encoder trained without ray-based regularization. Without ray-based regularization, the training of the auto-encoder quickly becomes unstable, resulting in severe floaters in the reconstructed mesh.

\subtitle{Quatitative ablation study on render-enhanced auto-encoder.}
To better assess the importance of incorporating render loss in our render-enhanced auto-encoder, we propose several variants and demonstrate the corresponding accuracy and volumeIoU on a validation set of Objaverse consisting of 2048 objects in \tref{tab:ablation}. Here, ``base'' represents the case with only BCE loss, while ``w/. 3D GAN loss'' represents incorporating the 3D patch-based GAN loss proposed in \citet{zheng2022sdfstylegan}.
\begin{table}[b]    
\vspace{-3mm}
\begin{center}
\caption{Quantitative ablation study on the proposed render-enhanced auto-encoder.}
\label{tab:ablation}
    \begin{tabular}{llcccccccccccccc}
        \toprule[2pt]
        Setting  & Accuracy$\uparrow$ & VolumeIoU$\uparrow$ \\
        \midrule
        w/o. $\mathcal{L}_{\text{normal}}^{\text{MSE}}$ & 95.972  & 89.977\\
        w/o. $\mathcal{L}_{\text{normal}}^{\text{LPIPS}}$ & 96.021 & 90.044\\
        w/. 3D patch GAN loss & 96.224 & 90.149 \\
        base & 94.745 & 87.164 \\
        \midrule
        Ours & \textbf{96.987} & \textbf{91.045} \\ 
        \bottomrule[2pt]
    \end{tabular}
    \end{center}
\end{table}

As shown in \tref{tab:ablation}, removing either the MSE loss or the LPIPS loss leads to a certain performance drop. Moreover, compared to the 3D patch-based GAN loss, the proposed render-based perceptual loss is more beneficial for auto-encoder training.

\subtitle{UV-space texture inpainting.}
In the right part of \fref{fig:more_ablations}, we compare the mesh obtained without UV inpainting. The figure clearly shows that without UV inpainting, colors may be missing from regions that are not visible from the fixed viewpoints of the multi-view diffusion. UV space inpainting effectively fills these regions, enhancing both the visual quality and realism of the model.

\subsection{More results}
\label{app:more_results}

\subtitle{Quantitative evaluation on texture generation.}
\begin{table}[!t]
    \centering
    \caption{\textbf{Quantitative comparisons on PBR texture generation.}
    $^\dagger$ denotes methods unable to generate PBR textures; baked-in textures are used as albedo for relighting.}
    \vspace{-3mm}
    \centering
    \renewcommand{\arraystretch}{0.5}
    \scalebox{0.8}{
    \begin{tabular}{l|ccccccccccccccc} 
    \toprule[2pt]
    \multirow{2}{*}{Method} & \multicolumn{2}{c}{Generative} & \multicolumn{2}{c}{Original lighting} & \multicolumn{2}{c}{Relighted} \\
    & FID$\downarrow$ & KID$\downarrow$ & PSNR$\uparrow$ & LPIPS$\downarrow$ & PSNR$\uparrow$ & LPIPS$\downarrow$ \\
    \midrule
    TEXTure & 28.03 & 7.6 & 18.74 & 0.157 & 16.21$^\dagger$ & 0.188$^\dagger$ \\
    Fantasia3D & 24.16 & 5.22 & 19.04 & 0.134 & 17.42 & 0.155 \\
    Paint3D & 25.28 & 5.19 & 19.2 & 0.144 & 16.04$^\dagger$ & 0.192$^\dagger$\\
    Ours & \textbf{20.14} & \textbf{3.21} & \textbf{22.87} & \textbf{0.089} & \textbf{22.44} & \textbf{0.098}\\
    \bottomrule[2pt]
    \end{tabular}}
    \vspace{-3.3mm}
    \label{tab:comp_texture}
\end{table}
The results in \tref{tab:comp_texture} show that MeshGen significantly outperforms other baselines across all metrics under both original and relighted settings, improving both reconstruction and generation metrics by a large margin.

\begin{figure*}[t]
    \centering
    \includegraphics[width=1.0\linewidth]{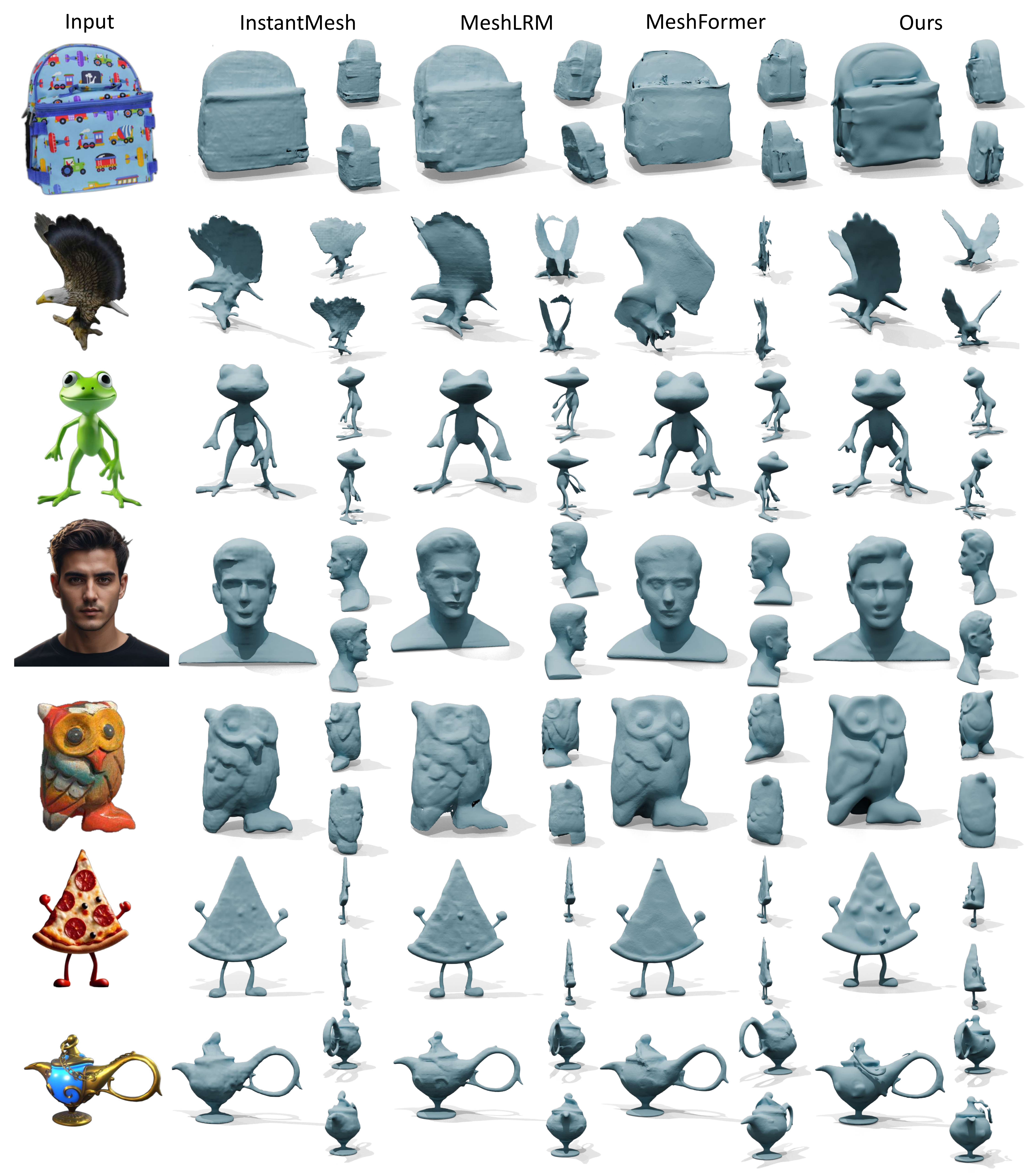}
    \caption{More comparisons with large reconstruction models, including InstantMesh~\citep{xu2024instantmesh}, MeshLRM~\citep{wei2024meshlrm}, MeshFormer~\citep{liu2024meshformer}.}
    \label{fig:more_comp_lrm}
\end{figure*}

\subtitle{More comparison with LRMs.}
In \fref{fig:more_comp_lrm}, we present more qualitative comparisons with large reconstruction models on image-to-shape generation, including InstantMesh~\citep{xu2024instantmesh}, MeshLRM~\citep{wei2024meshlrm}, and MeshFormer~\citep{liu2024meshformer}. Our method outperforms other large reconstruction models in capturing geometric details, such as the webbed feet of a frog and the handle of a backpack.

\begin{figure*}[t]
    \centering
    \includegraphics[width=1.0\linewidth]{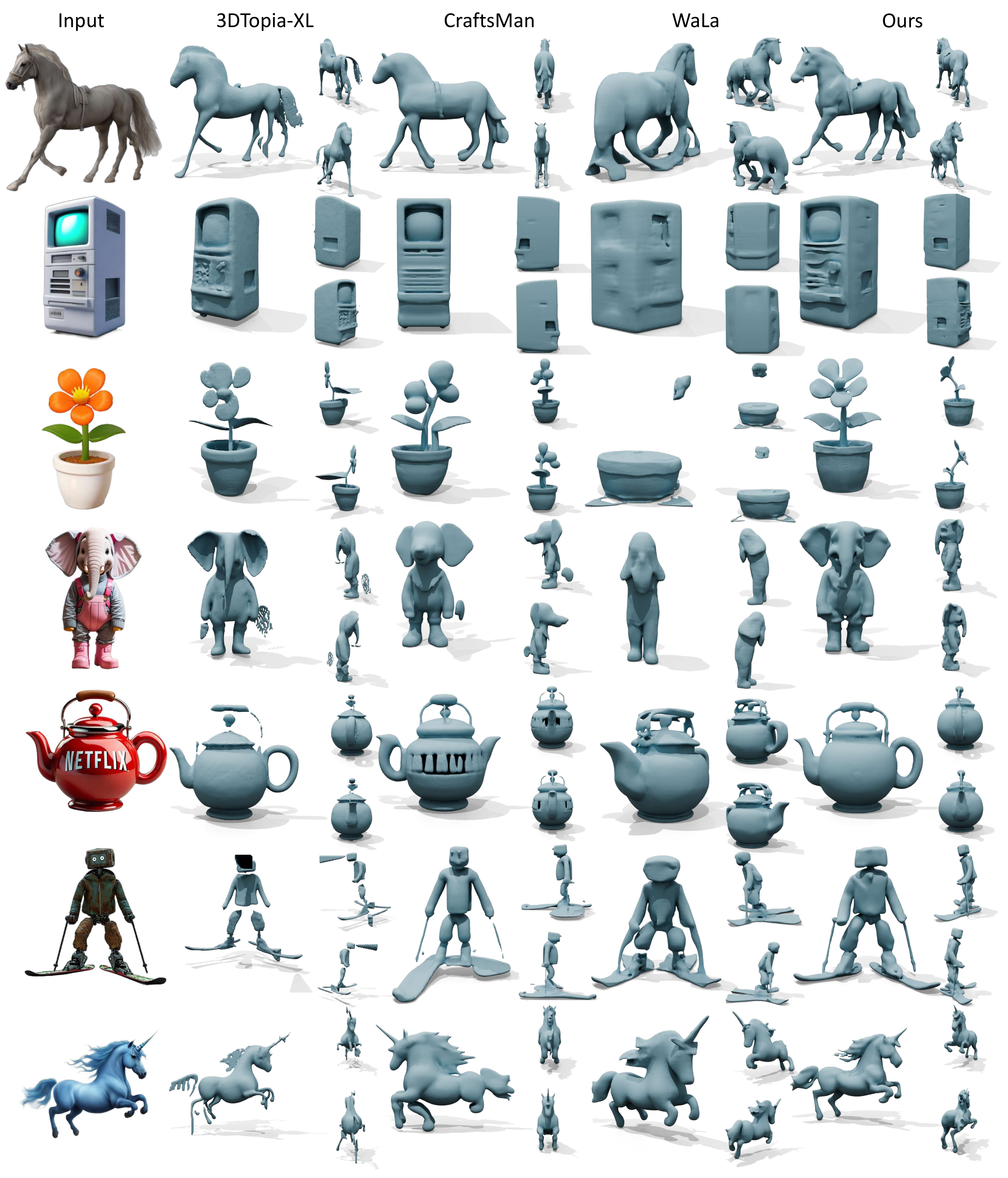}
    \caption{More comparisons with 3D native generation models, including, CraftsMan~\citep{li2024craftsman}, 3DTopia-XL~\citep{chen2024primx}, and WaLa~\citep{sanghi2024waveletlatentdiffusionwala}.}
    \label{fig:more_comp_native}
\end{figure*}

\subtitle{More comparison with 3D native methods.}
In \fref{fig:more_comp_native}, we present more qualitative comparisons with 3D native diffusion models, including CraftsMan~\citep{li2024craftsman}, 3DTopia-XL~\citep{chen2024primx}, and a more recent model with billion-level parameters, WaLa~\citep{sanghi2024waveletlatentdiffusionwala}. 
Clearly, the meshes generated by other methods are significantly lower in quality compared to ours and fail to produce meshes similar to the images, which introduces additional challenges for subsequent texturing. Although WaLa uses more parameters and a much larger training dataset than we do, thanks to our proposed augmentations, our method greatly outperforms WaLa in terms of mesh quality and image-shape alignment.

\subtitle{Compare with large reconstruction models with texture.}
\begin{figure*}[b]
    \centering
    \includegraphics[width=1.0\textwidth]{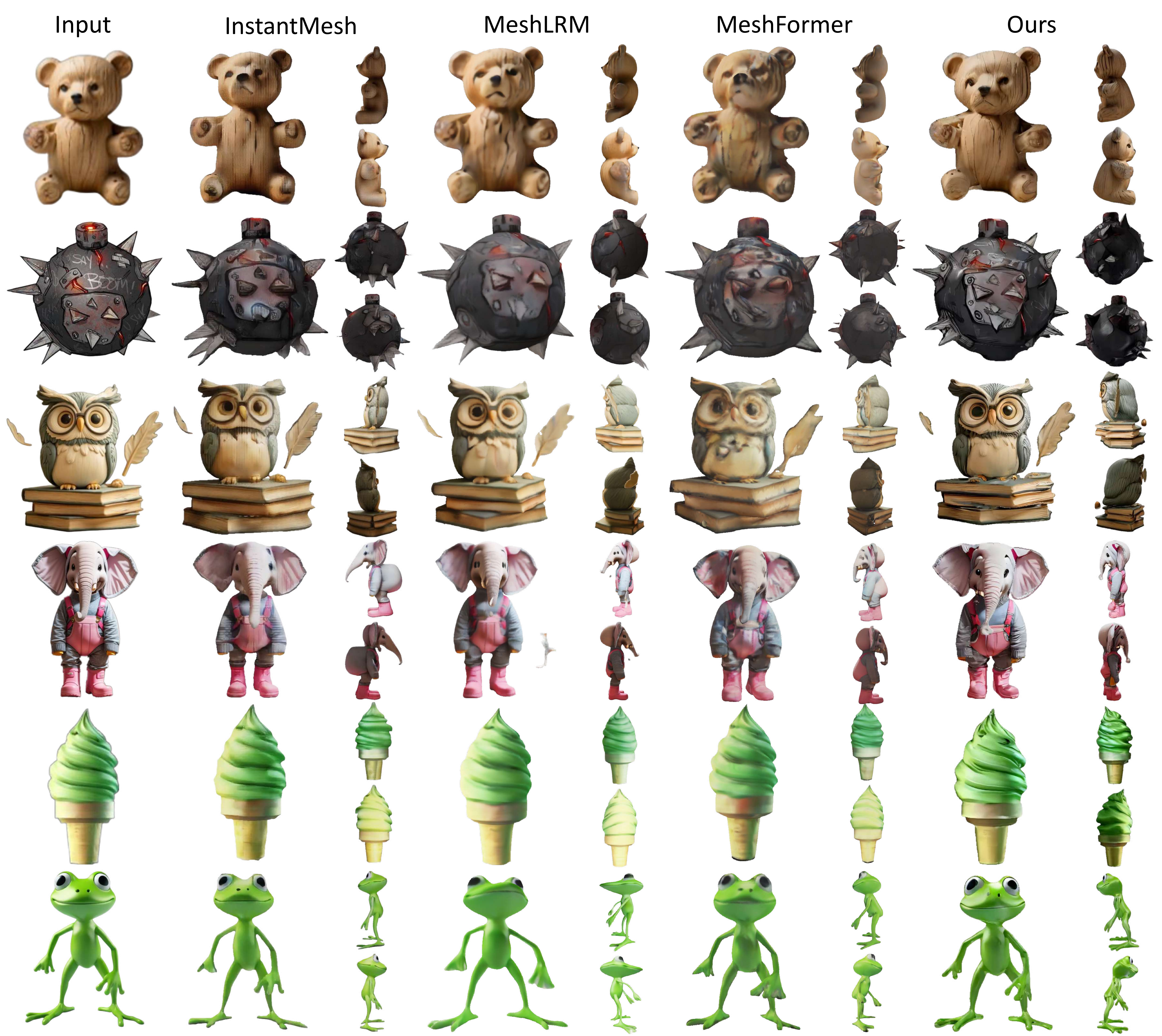}
    \caption{Qualitative comparison on textured meshes with state-of-the-art large reconstruction models, including InstantMesh~\citep{xu2024instantmesh}, MeshLRM~\citep{wei2024meshlrm} and MeshFormer~\citep{liu2024meshformer}.}
    \label{fig:comp_lrm_texture}
\end{figure*}
To comprehensively compare our approach with large reconstruction models, we compare the final generated textured mesh in \fref{fig:comp_lrm_texture}. It is evident from the figure that our method not only exceeds the previous best large reconstruction models in geometry but also produces clearer and more consistent textures.

\begin{figure*}[t]
    \vspace{-8mm}
    \centering
    \includegraphics[width=1.0\textwidth]{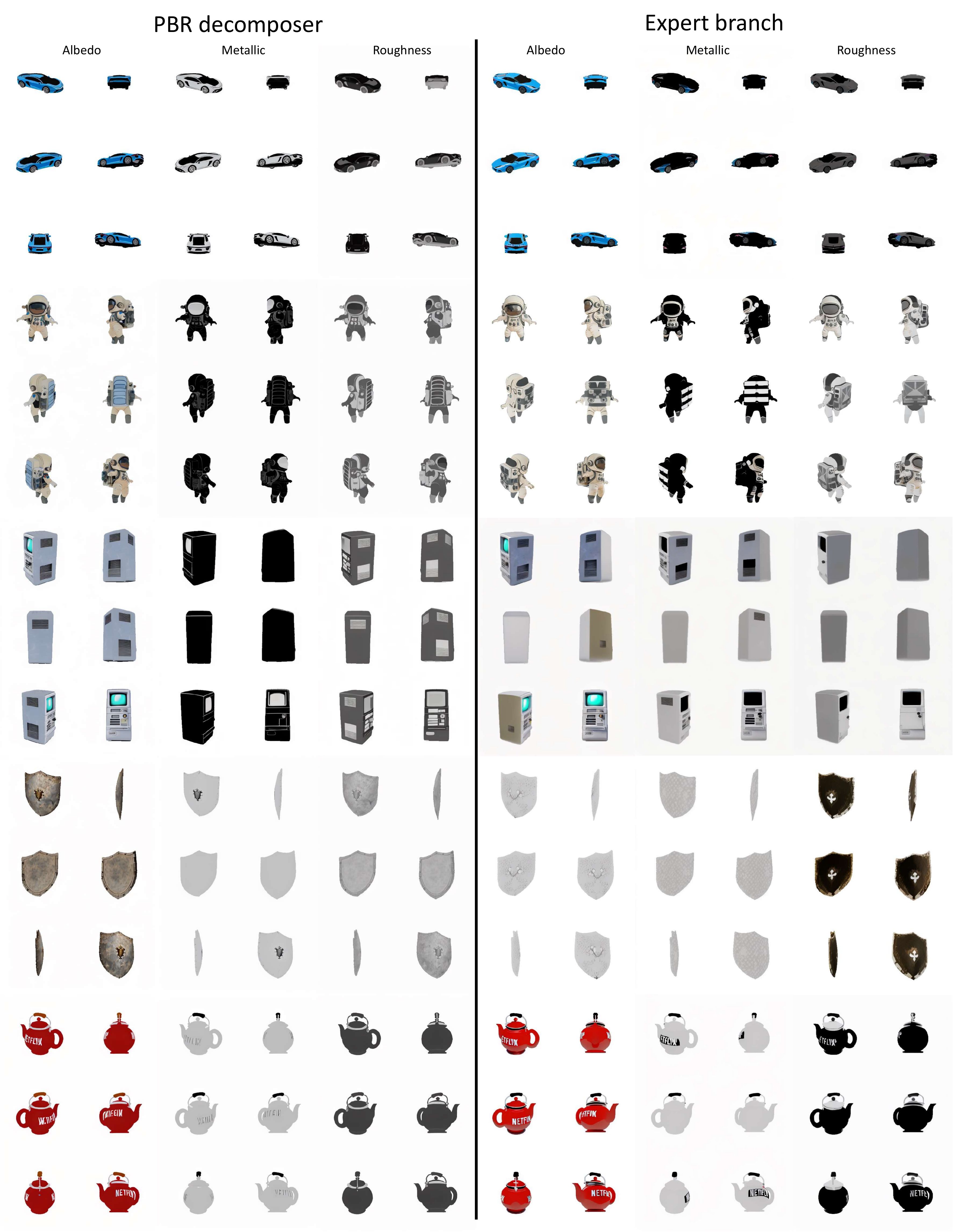}
    \caption{Comparison of PBR decomposer and expert branch on PBR component generation.}
    \label{fig:pbr_comp}
\end{figure*}

\subtitle{PBR joint generation v.s. PBR decomposition.}
In addition to our proposed multi-view PBR decomposer, previous methods like HyperHuman~\citep{hyperhuman} and CLAY~\citep{zhang2024clay} have suggested using expert branches to generate PBR components from image prompts.
Specifically, they use a shared backbone diffusion network and different input/output adapters to enable information sharing across different channels while exploiting information in pre-training.
To compare this approach with our proposed PBR decomposer, we trained a multi-view PBR component generator based on the expert branch approach using the same data. As shown in \fref{fig:pbr_comp}, we compare the PBR components generated by both methods for the same reference image and mesh. It can be seen that the model using the expert branch tends to produce PBR channels with color blending and inaccuracies, leading to unreasonable metallic and roughness outputs. We believe this is because different PBR channels should interact with the reference image in distinct ways during generation, and the shared parts of the expert branch hinder learning these different interactions. Additionally, compared to the expert branch-based model, our PBR decomposer inherently includes all multi-view information in the input, resulting in better consistency in the generated results. Furthermore, due to our limited data (or possibly because our method requires less data), we consider comparing both models with larger datasets as future work.

\begin{figure*}[t]
    \centering
    \vspace{-8mm}
    \includegraphics[width=0.85\linewidth]{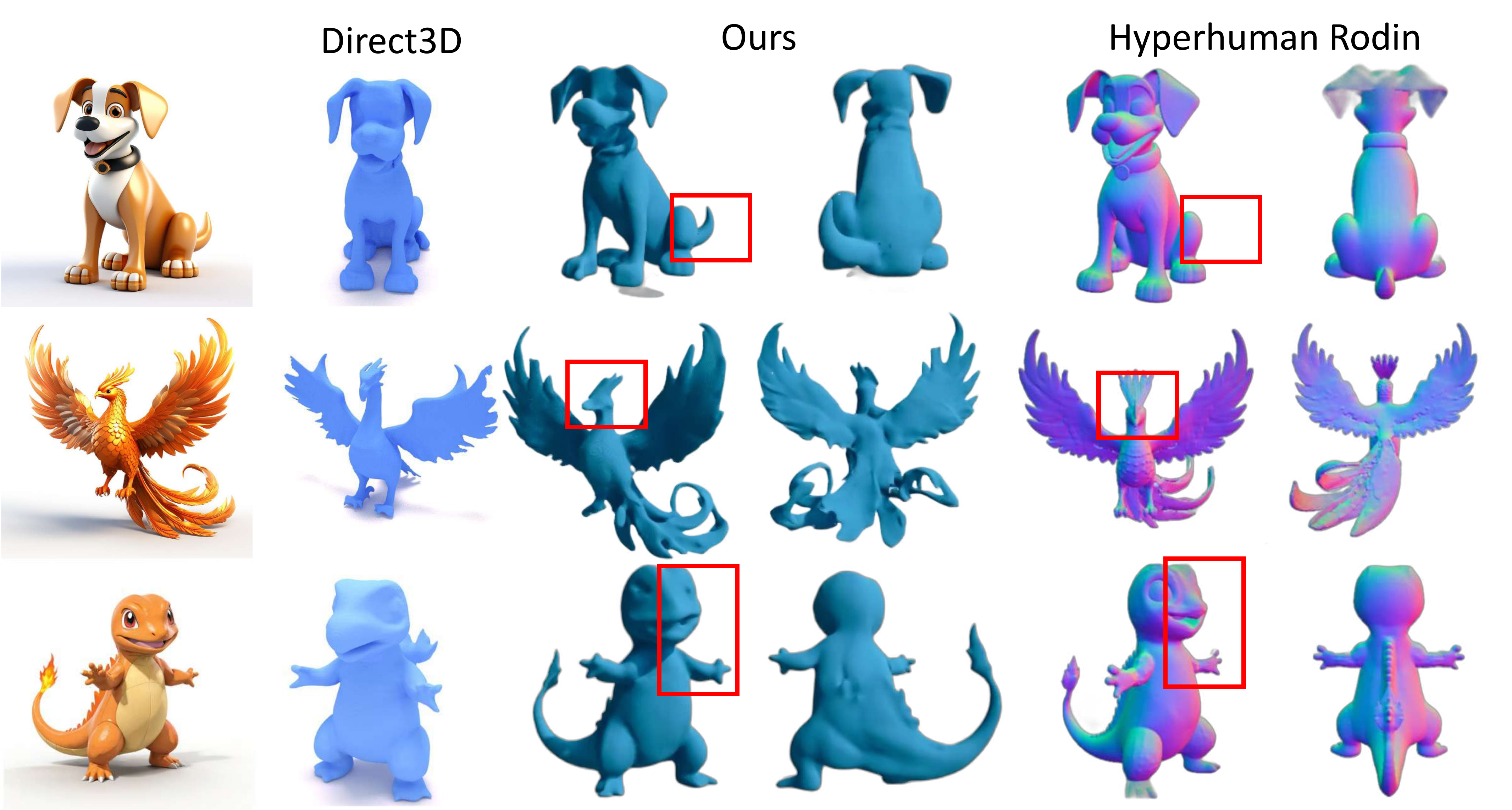}
    \caption{Comparison with non-open-source commercial products, including Direct3D and Hyperhuman Rodin.}
    \label{fig:comp_comm}
    \vspace{-5mm}
\end{figure*}

\subtitle{Compare with commercial products.}
In \fref{fig:comp_comm}, we compare our method with existing non-open-source commercial products. The results for Direct3D are sourced from their paper, while those for HyperHuman Rodin are generated on their official website without the ``symmetric'' tags. Although our method is currently limited by lacking high-quality data and computational resources, resulting in slightly lower mesh quality compared to commercial products, our proposed augmentation allows for better alignment with the images while other commercial products tend to generate symmetrical objects. We believe that with increased computational power and more high-quality data, our method can match the mesh quality of commercial products while preserving image-shape alignment.

\begin{figure*}[h]
    \centering
    \includegraphics[width=1.0\linewidth]{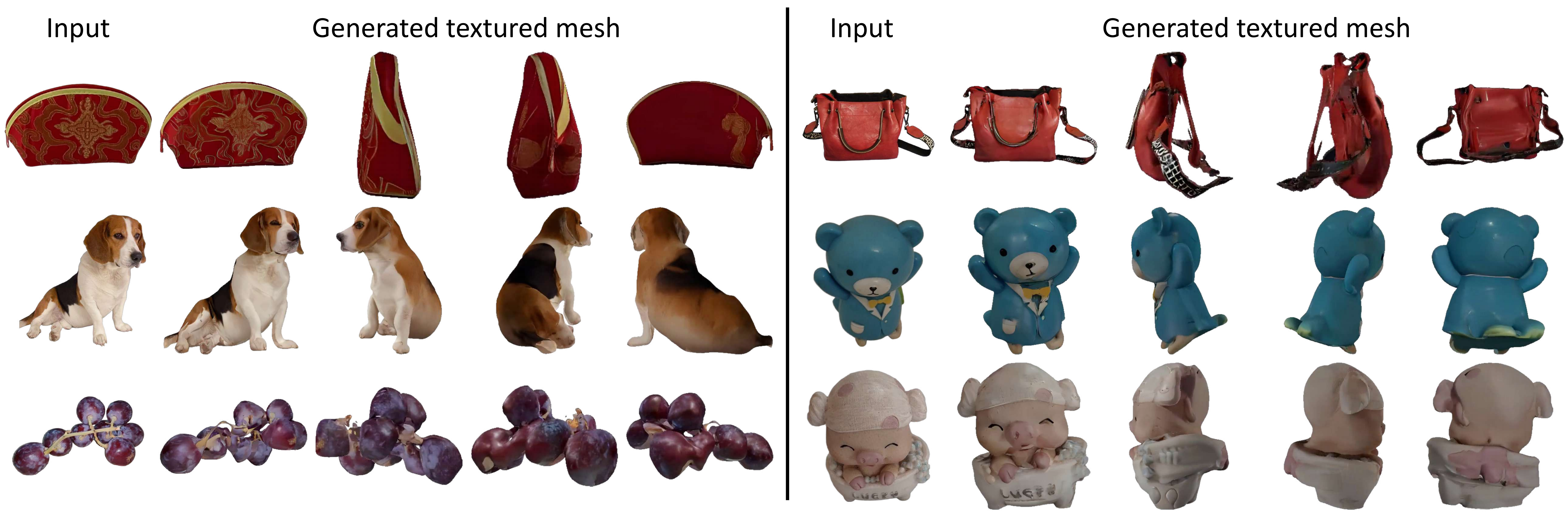}
    \caption{Performance of \approach on real-world captures.}
    \label{fig:ood}
\end{figure*}

\subtitle{Real-world images.}
To validate the performance of our method on real-world objects, we present a set of textured meshes generated from casual captures in \fref{fig:ood}. As shown in \fref{fig:ood}, our method is capable of generating reasonable shapes and consistent textures when processing real objects, demonstrating the generalization ability of our pipeline.

\begin{figure*}[t]
    \vspace{-8mm}
    \centering
    \includegraphics[width=0.9\textwidth]{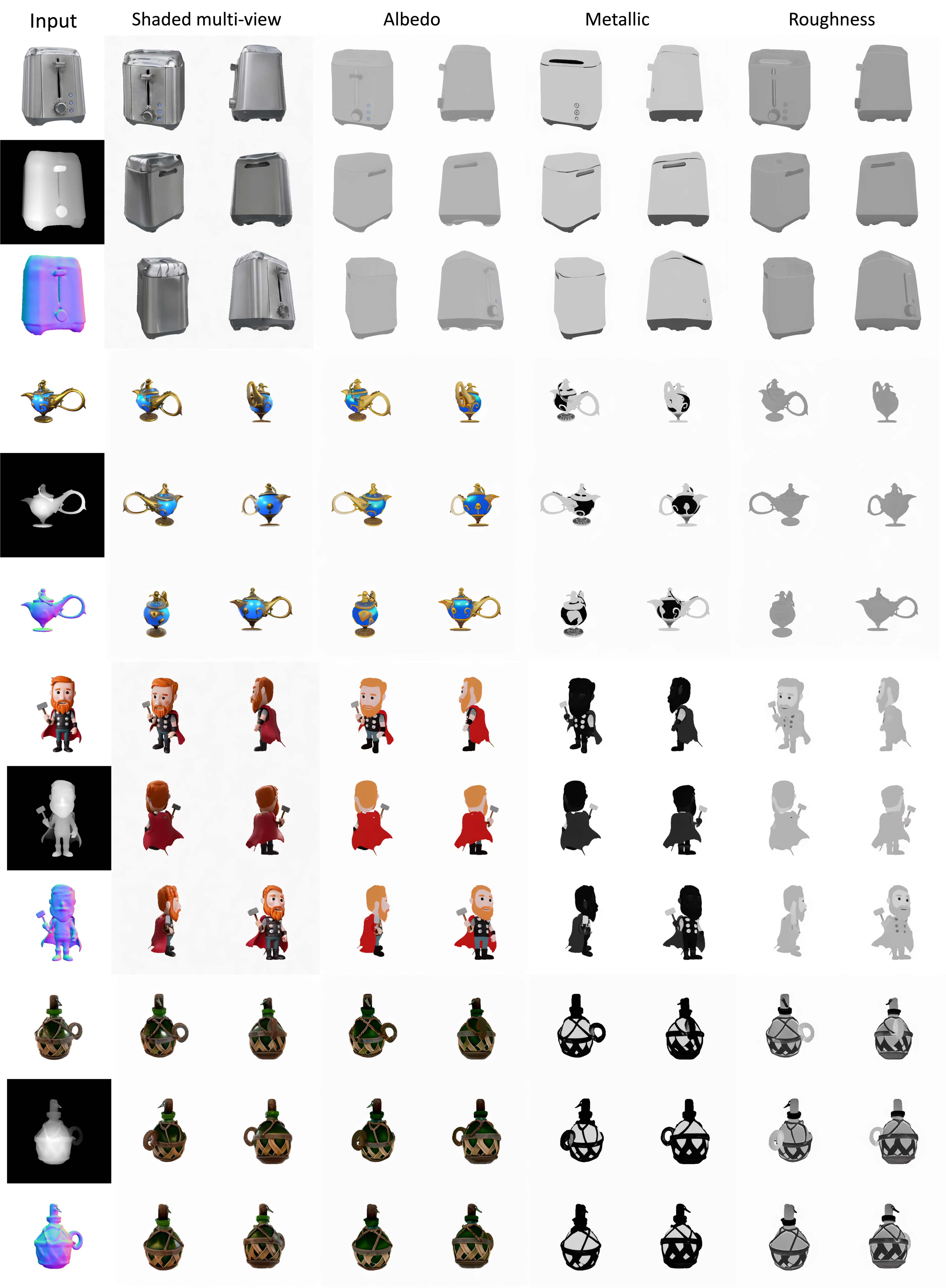}
    \caption{Intrinsic channels estimated using our multi-view PBR decomposer. The proposed PBR decomposer can handle images with complicated material under different lighting conditions.}
    \label{fig:pbr}
\end{figure*}

\subtitle{PBR decomposition results.}
In \fref{fig:pbr}, we present the intrinsic channels estimated using our proposed multi-view PBR decomposer. The results show that our PBR decomposer can accurately infer the PBR components of objects by leveraging multi-view information and can still generate multi-view consistent results under complex lighting conditions.

\section{Limitations}
\label{app:limitation}
\begin{figure*}[b]
    \centering
    \includegraphics[width=1.0\linewidth]{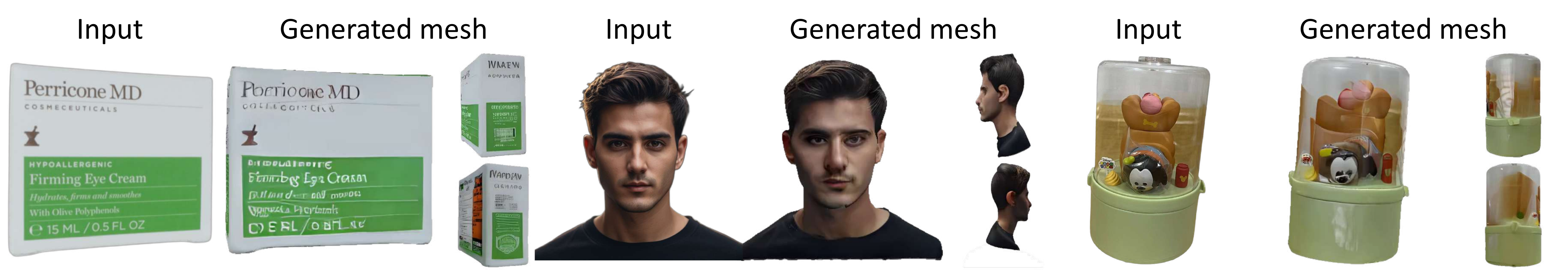}
    \caption{Some typical failure cases of \approach.}
    \label{fig:limitations}
\end{figure*}

Although our method has made some progress in native image-to-3D generation, there are still limitations in the following three areas.
\begin{enumerate}
    \item Due to the limited resolution of multi-view diffusion generation and the constraints of the auto-encoder used, our texture model struggles to accurately reproduce high-frequency details, such as the text on the box in the left part of \fref{fig:limitations}. We believe that using more advanced network architectures with more high-quality data could achieve higher-resolution multi-view generation.

    \item Our texture model finds it challenging to accurately capture textures and lighting effects from input images when dealing with objects with complex high-frequency information and lighting conditions, as shown by the face in the center of \fref{fig:limitations}.

    \item Our geometry and texture generation model currently cannot effectively handle transparent objects, as illustrated by the object on the right in \fref{fig:limitations}.
\end{enumerate}
Addressing these limitations will be the focus of our future research.

